# Dynamic phase transitions in thin ferromagnetic films

Hyunbum Jang[1], Malcolm J. Grimson[2], and Carol K. Hall[1]
[1]*Department of Chemical Engineering, North Carolina State University, Raleigh, NC 27695-7905*
[2]*Department of Physics, University of Auckland, Auckland, New Zealand*

Monte Carlo simulations have been used to investigate the dynamic phase behavior of a classical Heisenberg spin system with a bilinear exchange anisotropy $\Lambda$ in a planar thin film geometry. Studies of the field amplitude, frequency and temperature dependence show dynamic phase transitions in films subject to a pulsed oscillatory external field. Thin films with competing surface fields show separate and distinct dynamic phase transitions for the bulk and surface layers of the film. Between the two transitions, a mixed state with coexisting dynamically ordered and dynamically disordered phases is observed in the film. In contrast, the free film with no surface fields shows a single dynamic phase transition as in a bulk system.

## I. INTRODUCTION

The lag in the response of the magnetization for a ferromagnet subject to a time dependent oscillatory external driving field $H(t)$ is a well-known example of hysteresis. The delay is the result of a competition between the two time scales that characterize the non-equilibrium system, namely the period of the applied oscillatory field and the response time of the magnetization. In general, the form of the hysteresis curve for a ferromagnet is a function of the temperature as well as the amplitude and frequency of applied oscillatory external field. But of particular interest is the dynamic phase transition (DPT) observed in the hysteretic response of ferromagnets where the period averaged magnetization $Q$ passes from a dynamically ordered state with $|Q| > 0$ to a dynamically disordered state with $Q = 0$.

The frequency and field amplitude dependence of the hysteretic behavior in ferromagnets has been extensively studied both experimentally[1-4] and theoretically.[5-13] For the two-dimensional kinetic Ising model below its equilibrium critical temperature, a finite-size scaling analysis of large-scale Monte Carlo simulations has shown that the DPT is in the same universality class as the equilibrium Ising model.[14] A result confirmed in a recent study of a time-dependent Ginzburg-Landau model in an oscillatory field.[15]

The kinetic Ising model is a conceptually simple model with which to investigate the dynamic behavior of ferromagnets.[7-14] Furthermore it can provide a good representation of uniaxial ferromagnets in which magnetization reversal proceeds by nucleation and domain wall motion, it cannot account for magnetic relaxation processes such as the coherent rotation of spins. This requires a spin model with continuous degrees of freedom such as the classical Heisenberg model in which the magnetic spins can rotate through all possible orientations. The dynamic phase behavior of the anisotropic Heisenberg spin system in an applied sinusoidal oscillatory field was the subject of a recent study.[16] The inclusion of a bilinear exchange anisotropy $\Lambda$ in the model Hamiltonian gives the system Ising-like characteristics while allowing the magnetic spins to orient continuously. The DPT was studied as a function of the field amplitude, temperature, and frequency of the applied oscillatory field. For thin ferromagnetic films with competing surface fields, the form of the time-dependent layer magnetization across the film showed that dynamically ordered and dynamically disordered phases could coexist within the film with the dynamically disordered phase localized at one surface.

Finite size effects in thin films arising from both confinement and surface modification give rise to a variety of novel equilibrium phase behaviors that are not observed in the bulk materials. Of especial interest are thin films with competing surface fields, namely films with surface anisotropies in the direction perpendicular to the plane of the film that favor a positive magnetization at one surface and a negative magnetization at the other surface. Binder *et al*.[17-20] have made an extensive study of the thin ferromagnetic Ising film with competing surface forces and shown that the properties of the interface localization-delocalization transition are distinct from both the bulk ferromagnetic-paramagnetic phase transition and the wetting transition in semi-infinite systems. Complementary studies on thin ferromagnetic Heisenberg films with competing surface forces[21,22] have shown that the presence of an interface localization-delocalization transition is not restricted to discrete state models, but is also found in magnetic systems where the spins are continuously orientable, albeit with some degree of uniaxial anisotropy.

In this paper, we investigate the dynamic phase behavior of thin ferromagnetic films within the anisotropic Heisenberg model subject to a pulsed oscillatory external field. The inclusion of competing surface fields allows the magnetization distribution within the film to be controlled and its interplay with driving force provided by the applied oscillatory field to be studied. The temperature, field amplitude and frequency dependence of the dynamic response have been investigated by Monte Carlo simulation, extending and clarifying an earlier study of thin ferromagnetic films in a sinusoidal oscillatory field.[16] While the principal focus of this work is on films with competing surface fields, for comparative purposes the results for free films (without surface fields) and the corresponding bulk systems are also presented.

In the following section a full description of the model is given together with the details of the Monte Carlo simulation method. Section III contains the results of the simulations and is divided into five subsections. Subsection A focuses on the role of pulsed oscillatory external field amplitude, while the effects of temperature are presented in subsection B. The mixed state found in thin films with competing surface fields is investigated in subsection C. For the purposes of comparison the corresponding results for the free film with no surface fields and a bulk system are given in subsection D. The dependence of the dynamic phase behavior on the frequency of the applied oscillatory field is discussed in subsection E. The paper closes with a conclusion.

## II. MODEL

The Hamiltonian for the classical Heisenberg model with a bilinear exchange anisotropy $\Lambda$ can be written as[22]

$$\boldsymbol{H}_0 = -J \sum_{\langle i,j \rangle} \left( (1-\Lambda)\left(S_i^x S_j^x + S_i^y S_j^y\right) + S_i^z S_j^z \right) , \tag{1}$$

where $\mathbf{S}_i = (S_i^x,\ S_i^y,\ S_i^z)$ is a unit vector representing the $i$th spin and the notation $\langle i,j\rangle$ means that the sum is restricted to nearest-neighbor pairs of spins. $J$ is a coupling constant characterizing the magnitude of the exchange interaction and for ferromagnets $J > 0$. Following Binder and Landau,[23] $\Lambda$ determines the strength of the bilinear exchange anisotropy and is only applied to the $x$ and $y$ components of the spin. In the isotropic limit, $\Lambda = 0$, the model reduces to the familiar classical Heisenberg model, while for $\Lambda = 1$, the Hamiltonian becomes Ising-like.

The system under consideration here is a three dimensional thin planar film of finite thickness $D$ with Hamiltonian

$$\boldsymbol{H}(t) = \boldsymbol{H}_0 - \sum_{i \in \text{surface}\,1} \mathbf{H}_1 \cdot \mathbf{S}_i - \sum_{i \in \text{surface}\,D} \mathbf{H}_D \cdot \mathbf{S}_i - H(t) \sum_i S_i^z\ , \tag{2}$$

$\mathbf{H}_1$ and $\mathbf{H}_D$ are the static applied surface fields and the time dependent oscillatory external field $H(t)$ is taken to have a pulsed form with

$$H(t) = \begin{cases} -H_0\ , & \dfrac{2(k-1)\pi}{\omega} < t \le \dfrac{(2k-1)\pi}{\omega} \\ H_0\ , & \dfrac{(2k-1)\pi}{\omega} < t \le \dfrac{2k\pi}{\omega} \end{cases}\ , \tag{3}$$

where $H_0$ is the amplitude, $\omega$ is the angular frequency of the oscillatory external field, and $k$ ($k$ = 1,2,3,…) is an integer representing the number of periods of the pulsed oscillatory external field. The model film is a simple lattice of size $L \times L \times D$, in units of the lattice spacing. Periodic boundary conditions are applied in the $x$ and $y$ directions. Free boundary conditions are applied in the $z$ direction that is of finite thickness $D$. The system is subject to competing applied surface fields in layers $n = 1$ and $n = D$ of the film with

$$\mathbf{H}_1 = h\hat{\mathbf{z}}\delta_{i1}\ , \tag{4}$$
$$\mathbf{H}_D = -h\hat{\mathbf{z}}\delta_{iD}\ , \tag{5}$$

and the Hamiltonian for the system can be written as

$$\boldsymbol{H}(t) = \boldsymbol{H}_0 - h\left( \sum_{i \in \text{surface}\,1} S_i^z - \sum_{i \in \text{surface}\,D} S_i^z \right) - H(t) \sum_i S_i^z\ . \tag{6}$$

A film thickness $D = 12$ was used throughout. The value of $D = 12$ corresponds to the crossover regime between wall and bulk dominated behavior for thin Ising films.[18] In thinner films it is difficult to distinguish between "interface" and "bulk" phases in the film, since all layers of the film feel the effect of the competing surface fields rather strongly. While for thicker films the surfaces of the film only interact close to the bulk critical point. Unless otherwise stated, the results reported here are for lattices of size $L$ = 32. However no significant differences were found for lattices with $L$ = 64 and $L$ = 128 at non-critical values of $H_0$, $\omega$, and $T$. The Metropolis algorithm[24] was used in the Monte Carlo simulations with trial configurations were generated by the rotation of a randomly selected spin through a random angular displacement about one of the $x$, $y$, $z$ axes chosen at random.[25,26] A sequence of size $L \times L \times D$ trials comprises one Monte Carlo step per spin (MCSS), the unit of time in our simulations. The period of the pulsed oscillatory external field is given by product $R_{FS} \times N$, where $R_{FS}$ is the field sweep rate[27] and $N$ is a number of MCSS. The applied oscillatory field $H(t)$ being updated after every MCSS according to Eq. (3). Most simulations were performed for a value of $R_{FS} = 1$ with $N = 240$. In all of the simulations, the initial spin configuration was a ferromagnetically ordered state with $S_i = +1$ for all $i$ and $H(t = 0) = -H_0$ . Data from the initial cycles of the pulsed oscillatory field were discarded to avoid the effects of initial transients on the period averages of the measured quantities which were taken over a sequence of full cycles.

Most simulation studies of magnetization switching in ferromagnets have been based on the Monte Carlo method. The kinetic Ising model has been the focus of most of these studies, but vector spin models with continuous degrees of freedom have also been investigated.[28-31] Now while the Monte Carlo method is well established in the context of equilibrium systems, it must be noted that no physical time is associated with each trial. However, one MCSS corresponds to a series of random modifications of all the degrees of freedom of the system. Thus if the time rate by which a real system can modify all of its degrees of freedom is known by some independent argument, then the number of MCSS can be converted into a real time unit.[32,33]

The time-dependent magnetic order of the film is characterized by the $z$-component of the magnetization for the film

$$M_z(t) = \frac{1}{D}\sum_{n=1}^{D} M_n^z(t) \ , \qquad (7)$$

where

$$M_n^z(t) = \frac{1}{L^2}\sum S_i^z(t) \qquad (8)$$

is the time-dependent $z$-component of the magnetization for the $n$th layer of the film. The order parameter $Q$ for the DPT[10] is the period averaged magnetization over a complete cycle of the pulsed field defined by

$$Q = \frac{\omega}{2\pi}\oint M_z(t)\, dt \, . \qquad (9)$$

The period averaged magnetization for the $n$th layer of the film is given by

$$Q_n = \frac{\omega}{2\pi}\oint M_n^z(t)\, dt \, . \qquad (10)$$

The system exhibits a dynamically ordered phase with $|Q| > 0$ and a dynamically disordered phase with $Q = 0$. To provide further information on the location and nature of the DPT, fluctuations of the order parameter $\chi(Q)$ are measured in the simulations with

$$\chi(Q) = L^2 D \left( \left\langle Q^2 \right\rangle - \left\langle |Q| \right\rangle^2 \right) , \tag{11}$$

where $\langle\ \rangle$ denotes the average over a sequence of full cycles with initial transients discarded, and $L^2D$ is the number of spins in the system. Following Korniss *et al.*[14] the absolute order parameter, $|Q|$, is used in the definition of $\chi(Q)$ since in the dynamically ordered phase the probability density for $Q$ has peaks at both $+Q$ and $-Q$. Fluctuations in the order parameter for the $n$th layer of the film are quantified by

$$\chi(Q_n) = L^2 \left( \left\langle Q_n^2 \right\rangle - \left\langle |Q_n| \right\rangle^2 \right) . \tag{12}$$

Fluctuations in the energy of the film, $\chi(E)$, are monitored with

$$\chi(E) = L^2 D \left( \left\langle E^2 \right\rangle - \left\langle E \right\rangle^2 \right) , \tag{13}$$

where $E$ is the energy per spin for the system  Finite size scaling analyses of the DPT make use of Binder's fourth order cumulant. The fourth order cumulant for the period averaged magnetization, $U_L(Q)$, is defined as

$$U_L(Q) = 1 - \frac{\left\langle Q^4 \right\rangle}{3\left\langle Q^2 \right\rangle^2} . \tag{14}$$

Several groups have investigated the connection between the DPT and stochastic resonance (SR) in the kinetic Ising model driven by an oscillatory external field.[12,34,35] Most notably, double SR peaks have been observed,[12,34,35] one below and the other above the DPT. In a recent study, Kim *et al.*[35] have argued that the appearance of double SR peaks is a generic property of a system with a continuous DPT. Following Kim *et al.*[35] evidence for SR in the system under consideration here is obtained from measurement of the occupancy ratio $Q^{OR}$ defined by

$$Q^{OR} = \frac{\omega}{2\pi} \oint M_z(t) \frac{H(t)}{|H(t)|}\, dt , \tag{15}$$

where $H(t)/|H(t)|$ is the sign of the external pulsed oscillatory field.

## III. RESULTS

In this paper we have focused on a system with a bilinear exchange anisotropy of $\Lambda = 0.1$. For this weak exchange anisotropy the system is intermediate in character between the limiting Ising-like ($\Lambda = 1$) and Heisenberg ($\Lambda = 0$) models. In the absence of an applied oscillatory field, the bulk system displays a second order ferromagnetic-paramagnetic phase transition at a reduced temperature $T_c^* = k_B T_c / J = 1.53$, where $k_B$ is Boltzmann's constant. In the thin film geometry considered here with a film thickness $D$ = 12, the critical temperature characterizing the ferromagnetic-paramagnetic phase transition in the free film is lower than that for the corresponding bulk system with $T_c^* = 1.51$. When subject to competing surface fields with $h = -0.55$, the thin film with $D = 12$ exhibits an interface localization-delocalization phase transition with a critical temperature of $T_{ci}^* = 1.12$. This is well below the critical temperature of the ferromagnetic-paramagnetic phase transition for the bulk system and the free film.

### A. Field amplitude dependence of the DPT

The magnetization of the film becomes time dependent when it is subject to an applied oscillatory external field. The period averaged magnetization over a complete cycle of the pulsed oscillatory field, $Q$, characterizes the dynamic response of the film. First consider a film subject to competing surface fields with $h = -0.55$ and a pulsed oscillatory external field with field sweep rate $R_{FS} = 1$. This corresponds to a period for the pulsed oscillatory external field of 240 MCSS. Fig. 1 shows the mean period averaged magnetization, $\langle Q \rangle$, as a function of the pulsed oscillatory external field amplitude, $H_0$, at reduced temperatures of $T^* = 0.6$, 1.0, and 1.2. The quantity $\langle Q \rangle$ is determined from a sequence of full cycles with initial transients discarded. The error bars in the figure correspond to a standard deviation in the measured values and are only visible when they exceed the size of the symbol. The lines in the figure are only to guide the eye. At all three temperatures, in equilibrium with $H_0 = 0$, the systems are ferromagnetic. However, while the film displays a localized interface in the magnetization profile at the lower temperatures ($T^* = 0.6$ and 1.0), a delocalized interface is found at the highest temperature ($T^* = 1.2$). Note that the equilibrium interface localization-delocalization phase transition for a thin film with $D = 12$, $h = -0.55$, and $\Lambda = 0.1$ occurs at a critical temperature of $T_{ci}^* = 1.12$. It is immediately apparent from Fig. 1 that the qualitative form of $\langle Q \rangle$ as a function of $H_0$ at the lower temperatures $T^* = 0.6$ and 1.0 is different from that at the highest temperature $T^* = 1.2$. At temperatures $T^* = 0.6$ and 1.0, which are below the equilibrium interface localization-delocalization phase transition, the net magnetization of the film with competing surface fields is nonzero for $H_0 = 0$ and as a result $\langle Q \rangle \neq 0$ for $H_0 \to 0$. However at a temperature $T^* = 1.2$, which is above the interface localization-delocalization phase transition, the net magnetization of the film is zero when $H_0 = 0$ and so $\langle Q \rangle = 0$ as $H_0 \to 0$. At the lower $T^*$ values, $\langle Q \rangle$ is monotonically decreasing with increasing $H_0$. But at $T^* = 1.2$, $\langle Q \rangle$ initially rises with increasing $H_0$, before reaching a maximum and then decreasing to zero. The DPT is characterized by the vanishing of the order parameter $Q$ at a non-zero value of $H_0$. As $T^*$ increases, the location of the DPT shifts to lower values of $H_0$. At temperatures $T^* = 0.6$, 1.0, and 1.2, $\langle Q \rangle$ vanishes at values of $H_0 \approx 0.71$, 0.32, and 0.19, respectively. However it is difficult to locate the DPT directly due to the large fluctuations in $\langle Q \rangle$, particularly for the highest temperature.

Time series for the period averaged magnetization, $Q(t)$, at a temperature $T^* = 0.6$ are shown in Fig. 2 for pulsed oscillatory external fields of amplitude $H_0 = 0.3$, 0.7, and 2.0. For $H_0 = 0.3$, the film exhibits a dynamically ordered phase with $\langle Q \rangle \neq 0$, while for $H_0 = 2.0$, the system is dynamically disordered with $\langle Q \rangle = 0$. But in both cases the mean value of the period averaged magnetization is stable with only small fluctuations from the mean over the time series. In sharp contrast, the time series for $\langle Q \rangle$ at $H_0 = 0.7$ shows large fluctuations about a mean value $\langle Q \rangle \approx 0$ over the whole time series indicating that the system is near the DPT. Similar observations have been noted in studies of the DPT for the kinetic Ising model.[14]

The time dependence of the $z$-component of the magnetization, $M_z(t)$, at a temperature $T^* = 1.0$ is shown in Fig. 3 for a pulsed oscillatory external field of angular frequency $\omega = 2\pi / 240$ with amplitudes (a) $H_0 = 0.1$, (b) $H_0 = 0.3$, and (c) $H_0 = 1.0$. The figure shows the results of $M_z(t)$ for the film over five consecutive cycles of the pulsed oscillatory external field. Solid lines in the figure show the response curve $M_z(t)$, while the dotted lines correspond to the driving field $H(t)$. For $H_0 = 0.1$ in Fig. 3(a), the system is in a dynamically ordered state with $\langle Q \rangle \neq 0$. $M_z(t)$ has an oscillatory form with the same angular frequency as the pulse oscillatory external field $H(t)$. A qualitative difference in the form of $M_z(t)$ for sinusoidal[16] and pulsed oscillatory external fields is immediately apparent, particularly for small driving field amplitudes. For the sinusoidal external field, the $M_z(t)$ curve is also sinusoidal. However for the pulsed oscillatory external field, $M_z(t)$ has a saw-tooth form, while $H(t)$ is a square wave. In the pulsed oscillatory external field, following reversal of the applied field, there is an almost linear decay of the z-component of the magnetization toward its equilibrium value in the reversed applied field. If, as seen in Fig. 3(a), the decay of the magnetization from the metastable state is much slower than the frequency of the field reversal in the pulsed oscillatory external field, the period averaged magnetization will take a non-zero value. The almost linear decay of the magnetization from the metastable state following field reversal results in a saw-tooth form for $M_z(t)$. However, if the period of the pulsed oscillatory external field is comparable to decay time of the metastable magnetization state, as in Fig. 3(b), the period averaged magnetization tends to zero while the saw-tooth form for $M_z(t)$ is retained. Fig. 3(c) shows that when the decay time of the metastable magnetization state following field reversal is much shorter than the time between field reversals, the film has time to equilibrate after field reversal. Thus the form of the $M_z(t)$ curve is much closer to the square wave of the pulsed oscillatory external field.

More detailed information on the nature of the dynamic response of the film to the driving field is contained in Fig. 4. This shows the time-dependent layer magnetization across the film, $M_n^z(t)$, over three consecutive cycles of the applied oscillatory field with a frequency $f$ at a temperature $T^* = 1.0$ for (a) $H_0 = 0.1$ and (b) $H_0 = 1.0$. From Figs. 1 and 3, it can be seen that these correspond to systems (a) in the dynamically ordered state and (b) in the dynamically disordered state. For the film in a dynamically ordered state with $H_0 = 0.1$, $M_n^z(t)$ in Fig. 4(a) shows a large, almost constant, non-zero value for the mean magnetization over the time series at both surfaces, $n = 1$ and 12. But while $\langle M_n^z \rangle$ for $n = 1$ and 12 are approximately equal, they are opposite in sign. This indicates that the effects of the pulsed oscillatory external field for $H_0 = 0.1$ are much smaller than those of the competing surface fields with $|h| = 0.55$. The spins in the surface layers are ordered by the strong static surface fields and are not disturbed by the weak oscillatory external field. The interface between the regions of negative and positive magnetization of the film moves back and forth in response to the oscillatory external field due to coherent spin rotation of the spins in layers $n = 2$ - 5. The saw-tooth form of $M_z(t)$ as in Fig. 3(a) arises from the dynamic response of $M_n^z(t)$ in layers $n = 2$ - 5. The interface between the regions of positive and negative magnetization is located closer to layer $n = 1$ than layer $n = 12$ as a result of the initial conditions of the simulation. The positive value for $\langle M_z(t) \rangle$ in Fig. 3(a) is a result of the positive time-averaged layer magnetization of layers $n = 6 \sim 12$. For the film in a dynamically disordered state with $H_0 = 1.0$ shown in Fig. 4(b), $M_n^z(t)$ is almost uniform across the whole film. The uniform response of every layer of the film to the dominant oscillatory external field gives $\langle Q \rangle = 0$.

Fig. 5 shows the fluctuations of the order parameter, $\chi(Q)$, as a function of the pulsed oscillatory external field amplitude, $H_0$, for reduced temperatures of $T^* = 0.6$, 1.0, and 1.2. For $T^* = 0.6$ and 1.0, $\chi(Q)$ displays a characteristic single large peak. By comparison with the corresponding results of Fig. 1, the location of the peak in $\chi(Q)$ is seen to be close to, although just below, the DPT. However, for $T^* = 1.2$ no clear peak in $\chi(Q)$ is apparent. Instead, $\chi(Q)$ is seen to be large whenever $\langle Q \rangle \neq 0$. Further indication that the nature of the DPT differs between films with localized and delocalized interfaces in the equilibrium state.

### B. Temperature dependence of the DPT

The temperature dependence of the period averaged magnetization is shown in Fig. 6 for three amplitudes of the pulsed oscillatory external field: $H_0 = 0.3$, 0.55 and 1.0. Driving fields whose magnitude is below, equal to and above that of the size of the surface field $|h| = 0.55$. Error bars in the figure correspond to a standard deviation in the measured value and lines are only to guide the eye. Fig. 6 shows that for all three values of $H_{0,}$ the system exhibits a DPT between a dynamically ordered phase with $\langle Q \rangle \neq 0$ at low $T^*$ and a dynamically disordered with $\langle Q \rangle = 0$ at high $T^*$. However, $T_{cd}^*$, the critical temperature characterizing the DPT, decreases with increasing $H_0$ with $T_{cd}^* \approx 1.02$, 0.79, and 0.42 for the pulsed oscillatory external field amplitudes of $H_0 = 0.3$, 0.55, and 1.0, respectively. From the size of the error bars in the figure it is clear that there are large fluctuations in $\langle Q \rangle$ near the DPT.

Figure 7 shows the fluctuations in the order parameter, $\chi(Q)$, as a function of the temperature for pulsed oscillatory external field amplitudes: $H_0 = 0.3$, 0.55, and 1.0. In all cases there are distinct peaks in $\chi(Q)$ that are located at the temperatures close to, but just below, the DPT. Behavior also observed in Fig. 6.

### C. Mixed state

Large fluctuations in $Q$ close to the DPT arise from competition between the static surface fields and the pulsed oscillatory external field in the system. To isolate the surface effects, consider a surface order parameter for the film, $Q^{surface}$, defined by

$$Q^{surface} = \frac{1}{2}\left(Q_1 + Q_{12}\right) , \tag{16}$$

where $Q_1$ and $Q_{12}$ is the order parameter of the layers $n = 1$ and 12 of the film. Similarly define a bulk order parameter for the film, $Q^{bulk}$, with

$$Q^{bulk} = \frac{1}{2}\left(Q_6 + Q_7\right) . \tag{17}$$

$Q^{surface}$ and $Q^{bulk}$ are simply the mean of the period averaged magnetizations of the two surface and two central layers respectively. Fig. 8(a) shows $\langle Q^{surface} \rangle$ and $\langle Q^{bulk} \rangle$ as a function of $T^*$ for $H_0 = 0.3$. It is immediately clear from Fig. 8(a) that the central region and the surfaces of the film both show DPTs. But the critical temperature for the DPT in the surface layers, $T_{cd}^*$(surface), is not the same as $T_{cd}^*$(bulk), the critical temperature for the DPT in the bulk of the film. A comparison of the results in Figs. 6 and 8 shows that $T_{cd}^*$(surface) < $T_{cd}^*$, while $T_{cd}^*$(bulk) ≈ $T_{cd}^*$. Note however that the large fluctuations in $\langle Q \rangle$ and $\langle Q^{bulk} \rangle$ near the DPT make the accurate location of the DPT difficult in these cases.

The form of the temperature dependence of the bulk order parameter in Fig. 8(a) is qualitatively similar to that of the order parameter for the whole film in Fig. 6. But the region of large fluctuations in the order parameter close to the DPT is much smaller for the central region of the film than for the whole film. The temperature dependence of the surface order parameter is, however, significantly different. The DPT for the surface layers is much sharper and fluctuations in the surface order parameter in the vicinity of the DPT much smaller, a result of the pinning effects of the static surface fields. Fig. 8(a) suggests that the DPT for the film with competing surface fields is a composite of a series of DPTs for different regions of the film each of which may have a different critical temperature. Thus for certain temperatures one can expect some region of the film to be in a dynamically ordered state, while elsewhere there is a dynamically disordered state. We shall refer to such films as being in a mixed state.

Fig. 8(b) shows the temperature dependence of the order parameter for the *n*th layer, $Q_n$, across the whole film for $H_0$ = 0.3. At high temperatures, $T^* > T_{cd}^*$, the bulk of the film is in a dynamically disordered state with $Q$ = 0. Note that as a result of the competing surface fields, the $n$ = 1 surface layer has a negative non-zero period averaged magnetization, while the $n$ = 12 surface layer has a positive non-zero period averaged magnetization. But the contribution of the $n$ = 1 and $n$ = 12 layers to the surface dynamic order parameter are equal and opposite. So $Q^{surface}$ = 0 and the surfaces of the film can be regarded as being dynamically disordered even though $Q_1 \neq 0$ and $Q_{12} \neq 0$. The nonzero value of the period averaged magnetization for the $n$ = 1 and $n$ =12 layers is not the result of dynamic order in response to the oscillatory driving field, but arises from the coupling of the spins to the static surface field. As the temperature is reduced, the absolute value of the period averaged magnetization for the surface layers increases as a result of the reduction in thermal disorder. There is DPT in the bulk of the film at $T^* \sim 1.1$ and for $T^*$ < $T_{cd}^*$(bulk) the bulk of the film is dynamically ordered. However, note that for $T^*$ ~ $T_{cd}^*$(bulk), the period averaged magnetization of layers $n$ = 1 and $n$ = 12 remain equal and opposite. So the surface of the film can be said to remain dynamically disordered even though there is dynamic order in the bulk of the film. As the temperature is reduced further, the mixed state of the film persists until a DPT for the surface layers occurs at $T^*$ ~ 0.8. For temperatures below this the period averaged magnetizations of both the surface layer have the same sign. As a result there is sharp change in the surface dynamic order parameter for $T^* \approx T_{cd}^*$(surface). For $T^*$ < $T_{cd}^*$(surface) the applied surface fields are no longer able to maintain the coexistence of regions of positive and negative net magnetization within the film. As a result oscillatory driving field produces an almost uniform response of the whole film to the driving field. So only for temperatures $T^*$ < $T_{cd}^*$(surface) does the surface of the film become dynamically ordered, and only for $T^*$ < $T_{cd}^*$(surface) can the whole of the film be said to be in a dynamically ordered state. For $T_{cd}^*$(surface) < $T^*$ < $T_{cd}^*$(bulk) the film is in a mixed state where the dynamically ordered bulk of the film coexists with a dynamically disordered surface region.

Supplementary information on the mixed state and the bulk and surface DPTs in the thin film with competing surface fields is contained in Fig. 9. This shows the temperature dependence of (a) fluctuations of period averaged magnetization for the *n*th layer of the film, $\chi(Q_n)$, (b) the mean period averaged absolute value of the $z$-component magnetization for the film, $\langle|M_z|\rangle$, and (c) the mean period averaged occupancy ratio, $\langle Q^{OR}\rangle$, for $H_0$ = 0.3. Fig. 9(a) shows the fluctuations in the layer order parameter for the two surface layers $n$ = 1 and 12 and two central layers $n$ = 6 and 7. Large fluctuations in the period averaged magnetization for the surface layers are only seen for layer $n$ = 1 at temperatures corresponding the surface DPT. Whereas large fluctuations in the period averaged magnetization for the central region of the film are seen at temperatures in the mixed state. Of note in Fig. 9(b) is the sharp change $\langle|M_z|\rangle$ at the surface DPT resulting from the change in sign of the period averaged magnetization for layer $n$ = 1. Fig. 9(c) shows the mean period averaged occupancy ratio $\langle Q^{OR}\rangle$. The most remarkable feature of this figure is the *absence* of any significant features for temperatures corresponding to the bulk and surface DPTs.

The fourth order cumulant of the order parameter $U_L$ generally provides a strong indication of the nature of any underlying phase transition. For a continuous transition, $U_L$ decays monotonically from 2/3 to zero as the system moves form the ordered to the disordered phase. The intersection of $U_L$ as a function of the temperature for the various $L$ gives an estimate of the critical temperature. However for a first order transition $U_L$ develops a deep minimum whose location corresponds to the transition temperature. Fig. 10 shows the fourth order cumulant for the period averaged magnetization $U_L(Q)$, as a function of the temperature, $T^*$, for lattice sizes of $L$ = 16, 32, and 64 with $H_0$ = 0.3. The form of $U_L(Q)$ is markedly different from that of the bulk kinetic Ising model[14,36,37] and gives no evidence of a single DPT (either first order or continuous) for the film in the presence of competing surface fields.

### D. Bulk and free film

For comparative purposes, the dynamic phase transitions in both the bulk system and the free film subject to the same applied oscillatory field were investigated. Both being systems that have no surface fields ($h$ = 0). For the bulk system, periodic boundary conditions are applied in all directions. While for the free film the boundary conditions are the same as used for the film with competing surface fields, i.e. periodic boundary conditions in the $x$ and $y$ directions and free boundary conditions in the $z$ direction. Simulations for both systems used a 32 × 32 ×12 simple cubic lattice, the same type and size of lattice as used in simulations of the film with competing surface fields.

The temperature dependence of the period averaged magnetization is shown in Fig. 11 for three amplitudes of the pulsed oscillatory external field: $H_0$ = 0.3, 0.55, and 1.0. The qualitative form of $\langle Q\rangle$ as a function of $T^*$ is the same for both the bulk system and the free film. At low $T^*$, $\langle Q\rangle \neq 0$ and the system is dynamically ordered, while a dynamically disordered state is found at high $T^*$ with $\langle Q\rangle = 0$. The critical temperature characterizing the DPT, $T_{cd}^*$, is seen to be slightly lower for the free film than for the bulk system. A direct comparison of Fig.11 with the results in Fig. 6 shows that for all three amplitudes of the pulsed oscillatory external field the DPT is much sharper for the free film and the bulk system. It also shows that $T_{cd}^*$ is much smaller for the film with competing surface fields.

Supplementary information on the DPT is presented in Fig. 12. For $H_0$ = 0.3 this shows: (a) the fluctuations of the dynamic order parameter, $\chi(Q)$, (b) the fluctuations of the energy, $\chi(E)$, and (c) the mean period averaged occupancy ratio, $\langle Q^{OR}\rangle$, as a function of the temperature, $T^*$. Peaks in $\chi(Q)$ and $\chi(E)$ close to the DPT are immediately apparent.

In Fig. 12(c) $\langle Q^{OR} \rangle$ shows double stochastic resonance peaks, one above and the other below the DPT. Behavior seen elsewhere in similar systems.[34,35] Fig. 12(c) indicates that the DPT for the bulk system occurs at $T^* \approx 1.30$, while the DPT for the free film is located at $T^* \approx 1.28$. These estimates of the critical temperatures for the bulk and free film are consistent with the temperatures at which $\langle Q \rangle$ vanishes in Fig. 11. It is interesting to note that no evidence of stochastic resonance at the DPT was observed for the thin film with competing surface fields in Fig. 9(c).

Finite size scaling results for the bulk system are shown in Fig. 13 for lattice sizes of $L = 16$, 32, and 64 with a pulsed oscillatory external field of amplitude $H_0 = 0.3$. The form of $U_L(Q)$ is consistent with a continuous DPT and the intersection of $U_L(Q)$ for the various $L$ is located at $T^* \approx 1.29$. Close to other estimates for the location of the DPT from Figs. 11 and 12.

### E. Frequency dependence

The dependence of the DPT in the film with competing surface fields ($h = -0.55$) on the frequency of the applied oscillatory field is presented in Fig. 14. This shows the period averaged magnetization, $Q$, as a function of field sweep rate, $R_{FS}$, at a temperature $T^* = 0.6$ for pulsed oscillatory external fields of amplitude $H_0 = 0.3$ and 0.55. The period of the pulsed oscillatory field is set to $240 \times R_{FS}$ MCSS. Only integer values of $R_{FS}$ are considered. At all values of $H_0$, the film displays a dynamically ordered phase with $\langle Q \rangle \neq 0$ for small $R_{FS}$ (high frequency oscillatory external fields), while at large $R_{FS}$ (low frequency oscillatory external fields) a dynamically disordered phase with $\langle Q \rangle = 0$ is found. The critical frequency characterizing the DPT for the film decreases with decreasing $H_0$. Since this work uses a discrete set of frequencies for the driving field, only the general features of the frequency dependence of the DPT are found. More details of the frequency dependence of the DPT in the kinetic Ising model are given elsewhere.[14]

Fig. 15 shows results for the fluctuations in the film near the DPT. For $H_0 = 0.3$ and $T^* = 0.6$, the figure shows the dependence on the field sweep rate $R_{FS}$ of (a) the fluctuations of the order parameter, $\chi(Q)$, (b) the fluctuations of the energy, $\chi(E)$, and (c) the mean period averaged occupancy ratio, $\langle Q^{OR} \rangle$. It is immediately clear that $\chi(Q)$ and $\chi(E)$ both show broad peaks at $R_{FS} = 7$, while $\langle Q^{OR} \rangle$ has a minimum at $R_{FS} = 7$. This is the value for $R_{FS}$ at which $\langle Q \rangle$ vanishes in Fig. 14. Thus the critical frequency characterizing the DPT is located around $R_{FS} = 7$. No evidence for the mixed state was observed for the discrete set of field frequencies used in this work. Notably in Fig. 15(c) $\langle Q^{OR} \rangle$ yields a local minimum at the DPT similar to results for the bulk and free film systems seen in Fig. 12(c).

## IV. Conclusion

The dynamic response of thin ferromagnetic Heisenberg films with competing surface fields to a pulsed oscillatory external field has been studied. The magnetic spins in the model are continuously orientable, but the bilinear exchange anisotropy $\Lambda$ in the Heisenberg Hamiltonian ensures that Ising-like characteristics are retained. At low temperatures the competition between the ferromagnetic ordering tendencies of the spins and the applied oscillatory field determines the behavior of the film, which exhibits a dynamic phase transition between dynamically ordered and dynamically disordered phases.

The bulk and free film show a DPT where the system moves from a dynamically disordered phase with $Q = 0$ to a dynamically ordered state where $|Q|$ is non-zero. This dynamically ordered state can be of either positive or negative net magnetization, i.e. either $+Q$ or $-Q$, the one selected depending on the initial conditions and random number sequence of the simulation. While the system can fluctuate between the positive and negative magnetization states, the two states will not coexist at the same time in a small system. As a result the DPT is associated with stochastic resonance in the system.

In the film with competing surface fields, the dynamic response of the film is markedly different from the free film. A DPT is observed, but at a lower temperature or field amplitude, and fluctuations of the order parameter in the vicinity of the DPT are much greater. These are a result of the interplay of the static surface fields with ferromagnetic ordering of the spins and the driving force of the oscillatory external field. The competing surface fields pin the magnetization of the surface layers and at low temperatures localize the interface between regions of positive and negative magnetization near one surface of the film. In the presence of an external driving field, this leads to a coexistence of dynamically ordered and dynamically disordered states within the film. As a result the DPT occurs at different temperatures in different regions of the film. Only at very low temperatures does the dynamic response of the whole film become uniform. Thus the DPT for the film with competing surface fields is a composite of DPTs for different regions of the film spread over a range of temperature. Finite-size scaling results for the fourth order cumulant for the period averaged magnetization $U_L(Q)$ show that for the films with competing surface fields there is no evidence of a single DPT (either first order or continuous) in the film. In contrast, finite-size scaling results for $U_L(Q)$ give us a clear evidence of the DPT in the cumulant crossing for the bulk system and the free film. Furthermore there is no evidence for stochastic resonance in the film with competing surfaces. This is in marked contrast to the observation of double stochastic resonance peaks about the DPT for the free film and corresponding bulk system.

**References**

[1]Y.-L. He and G.-C. Wang, Phys. Rev. Lett. **70**, 2336 (1993).
[2]Q. Jiang, H.-N. Yang, and G.-C. Wang, Phys. Rev. B **52**, 14911 (1995).
[3]Jih-Shin Suen and J. L. Erskine, Phys. Rev. Lett. **78**, 3567 (1997).
[4]Jih-Shin Suen, M. H. Lee, G. Teeter, and J. L. Erskine, Phys. Rev. B **59**, 4249 (1999).
[5]T. Tomé and M. J. de Oliveira, Phys. Rev. A **41**, 4251 (1990).
[6]C. N. Luse and A. Zangwill, Phys. Rev. E **50**, 224 (1994).
[7]W. S. Lo and R. A. Pelcovits, Phys.Rev. A **42**, 7471 (1990).
[8]M. Acharyya and B. K. Chakrabarti, Phys.Rev. B **52**, 6550 (1995)
[9]M. Acharyya, Phys. Rev. E **58**, 179 (1998).
[10]B. K. Chakrabarti and M. Acharyya, Rev. Mod. Phy. **71**, 847 (1999).
[11]S. W. Sides and P. A. Rikvold, J. Appl. Phys. **83**, 6494 (1998).
[12]S. W. Sides, P. A. Rikvold, and M. A. Novotny, Phys. Rev. E **57**, 6512 (1998).
[13]S. W. Sides, P. A. Rikvold, and M. A. Novotny, Phys. Rev. E **59**, 2710 (1999).
[14]G. Korniss, C. J. White, P. A. Rikvold, and M. A. Novotny, Phys. Rev. E **63**, 016120 (2001).
[15]H. Fujisaka, H. Tutu, and P. A. Rikvold, Phys. Rev. E **63**, 036109 (2001).
[16]H. Jang and M. J. Grimson, Phys. Rev. E **63**, 066119 (2001).
[17]K. Binder, D. P. Landau, and A. M. Ferrenberg, Phys. Rev. Lett. **74**, 298 (1995).
[18]K. Binder, D. P. Landau, and A. M. Ferrenberg, Phys. Rev. E **51**, 2823 (1995).
[19]K. Binder, R. Evans, D. P. Landau, and A. M. Ferrenberg, Phys. Rev. E **53**, 5023 (1996).
[20]A. M. Ferrenberg, D. P. Landau and K. Binder, Phys. Rev. E **58**, 3353 (1998).
[21]H. Jang and M. J. Grimson, Phys. Rev. B **55**, 12556 (1997).
[22]H. Jang and M. J. Grimson, J. Phys: Condens. Matter **10**, 9641 (1998).
[23]K. Binder and D. P. Landau, Phys. Rev. B **13**, 1140 (1976).
[24]N. Metropolis, A.W. Rosenbluth, M. N. Rosenbluth, A. H. Teller, and E. Teller, J. Chem. Phys. **21**, 1087 (1953).
[25]R. E. Watson, M. Blume, and G. H. Vineyard, Phys. Rev. **181**, 811 (1969).
[26]J. R. Barker and R. O. Watts, Chem. Phys. Lett. **3**, 144 (1969).
[27]L. C. Sampio, M. P. de Albuquerque and F. S. de Menezes, Phys. Rev. B **54**, 6465 (1996).
[28]J. M. Gonzalez, R. Ramirez, R. Smirnov-Rueda and J. Gonzalez, Phys. Rev. B **52**, 16034 (1995).
[29]J. M. Gonzalez, R. Smirnov-Rueda and J. Gonzalez, J. Appl. Phys. **81**, 5573 (1997).
[30]D. Hinzke and U. Nowak, Phys. Rev. B **58**, 265 (1998).
[31]U. Nowak and D. Hinzke, J. Appl. Phys. **85**, 4337 (1999).
[32]U. Nowak, R. W. Chantrell and E. C. Kennedy, Phys. Rev. Lett. **84**, 163 (1999).
[33]D. Hinzke and U. Nowak, Phys. Rev. B **61**, 6734 (2000).
[34]K-t. Leung and Z. Neda, Phys. Lett. A **246**, 505 (1998).
[35]B. J. Kim, P. Minnhagen, H. J. Kim, M. Y. Choi, and G. S. Jeon, Europhys. Lett. 56, 333 (2001).
[36]S. W. Sides, P. A. Rikvold, and M. A. Novotny, Phys. Rev. Lett. **81**, 834 (1998).
[37]M. Acharyya, Phys. Rev. E **59**, 218 (1999).

## Figure Captions

**Fig. 1** Period averaged magnetization, $\langle Q \rangle$, as a function of the pulsed oscillatory external field amplitude of applied, $H_0$, for temperatures $T^*$ = 0.6, 1.0, and 1.2.
**Fig. 2** Time series of the period averaged magnetization, $Q(t)$, at a temperature $T^*$ = 0.6 for pulsed oscillatory external fields of amplitude $H_0$ = 0.3, 0.7, and 2.0.
**Fig. 3** Dynamic response of the $z$-component of the magnetization for the film, $M_z(t)$, at a temperature $T^*$ = 1.0 to a pulsed oscillatory external field with amplitude: (a) $H_0$ = 0.1, (b) $H_0$ = 0.3, and (c) $H_0$ = 1.0.
**Fig. 4** Dynamic response of the layer magnetization across the film, $M_n^z(t)$, at a temperature $T^*$ = 1.0 for pulsed oscillatory external field amplitudes, (a) $H_0$ = 0.1 and (b) $H_0$ = 1.0.
**Fig. 5** Fluctuations of the order parameter, $\chi(Q)$, as a function of the amplitude of pulsed oscillatory external field, $H_0$, for temperatures $T^*$ = 0.6, 1.0, and 1.2.
**Fig. 6** Period averaged magnetization, $\langle Q \rangle$, as a function of the temperature, $T^*$, for pulsed oscillatory external field amplitudes of $H_0$ = 0.3, 0.55, and 1.0.
**Fig. 7** Fluctuations of the order parameter, $\chi(Q)$, as a function of the temperature, $T^*$, for pulsed oscillatory external field amplitudes of $H_0$ = 0.3, 0.55, and 1.0.
**Fig. 8** (a) Surface order parameter, $\langle Q^{surface} \rangle$, and bulk order parameter, $\langle Q^{bulk} \rangle$, for the film, and (b) period averaged magnetization for the $n$th layer of the film, $\langle Q_n \rangle$, as a function of the temperature, $T^*$, for a pulsed oscillatory external field amplitude of $H_0$ = 0.3.
**Fig. 9** (a) Fluctuations of the period averaged magnetization of the $n$th layer, $\langle \chi(Q_n) \rangle$, (b) period averaged absolute value of the $z$-component magnetization of the film, $\langle |M_z| \rangle$, (c) period averaged occupancy ratio, $\langle Q^{OR} \rangle$, as a function of the temperature, $T^*$, for a pulsed oscillatory external field amplitude $H_0$ = 0.3.
**Fig. 10** Fourth order cumulant for the period averaged magnetization, $U_L(Q)$, as a function of the temperature, $T^*$, with a pulsed oscillatory external field amplitude $H_0$ = 0.3 for lattice sizes of $L$ = 16, 32, and 64.
**Fig. 11** Period averaged magnetization, $\langle Q \rangle$, as a function of the temperature, $T^*$, for pulsed oscillatory external field amplitudes of $H_0$ = 0.3, 0.55, and 1.0 in the free film with no surface fields (open symbols) and the bulk system (solid symbols).
**Fig. 12** (a) Fluctuations of the order parameter, $\chi(Q)$, (b) fluctuations of the energy, $\chi(E)$, and (c) period averaged occupancy ratio, $\langle Q^{OR} \rangle$, as a function of the temperature, $T^*$, for a pulsed oscillatory external field amplitude $H_0$ = 0.3 in the free film with no surface fields (open symbols) and the bulk system (solid symbols).
**Fig. 13** Fourth order cumulant for the period averaged magnetization, $U_L(Q)$, as a function of the temperature, $T^*$, with a pulsed oscillatory external field amplitude $H_0$ = 0.3 for lattice sizes of $L$ = 16, 32, and 64 in the bulk system. Inset is an enlargement of the cumulant crossing region.
**Fig. 14** Period averaged magnetization, $\langle Q \rangle$, as a function of the field sweep rate, $R_{FS}$, at a temperature $T^*$ = 0.6 for pulsed oscillatory external fields of amplitude $H_0$ = 0.3 and 0.55.
**Fig. 15** (a) Fluctuations of the order parameter, $\chi(Q)$, (b) fluctuations of the energy, $\chi(E)$, and (c) period averaged occupancy ratio, $\langle Q^{OR} \rangle$, as a function of the field sweep rate, $R_{FS}$, at a temperature $T^*$ = 0.6 for a pulsed oscillatory external field amplitude $H_0$ = 0.3.

Fig. 1

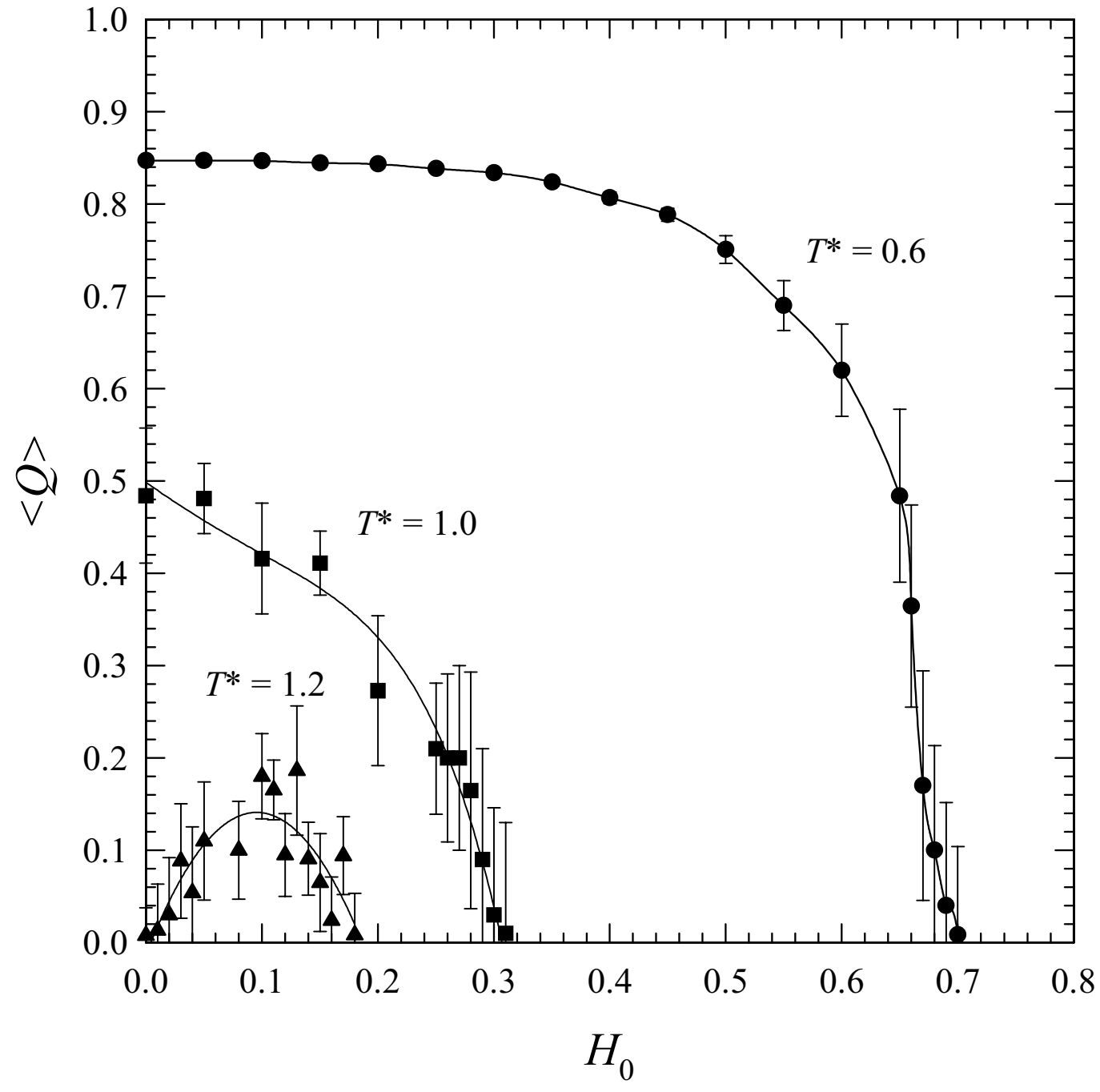

1.0
0.9
0.8
0.7
0.6
0.5
0.4
0.3
0.2
0.1
0.0
<Q>
T* = 0.6
T* = 1.0
T* = 1.2
0.0 0.1 0.2 0.3 0.4 0.5 0.6 0.7 0.8
H0

Fig. 2

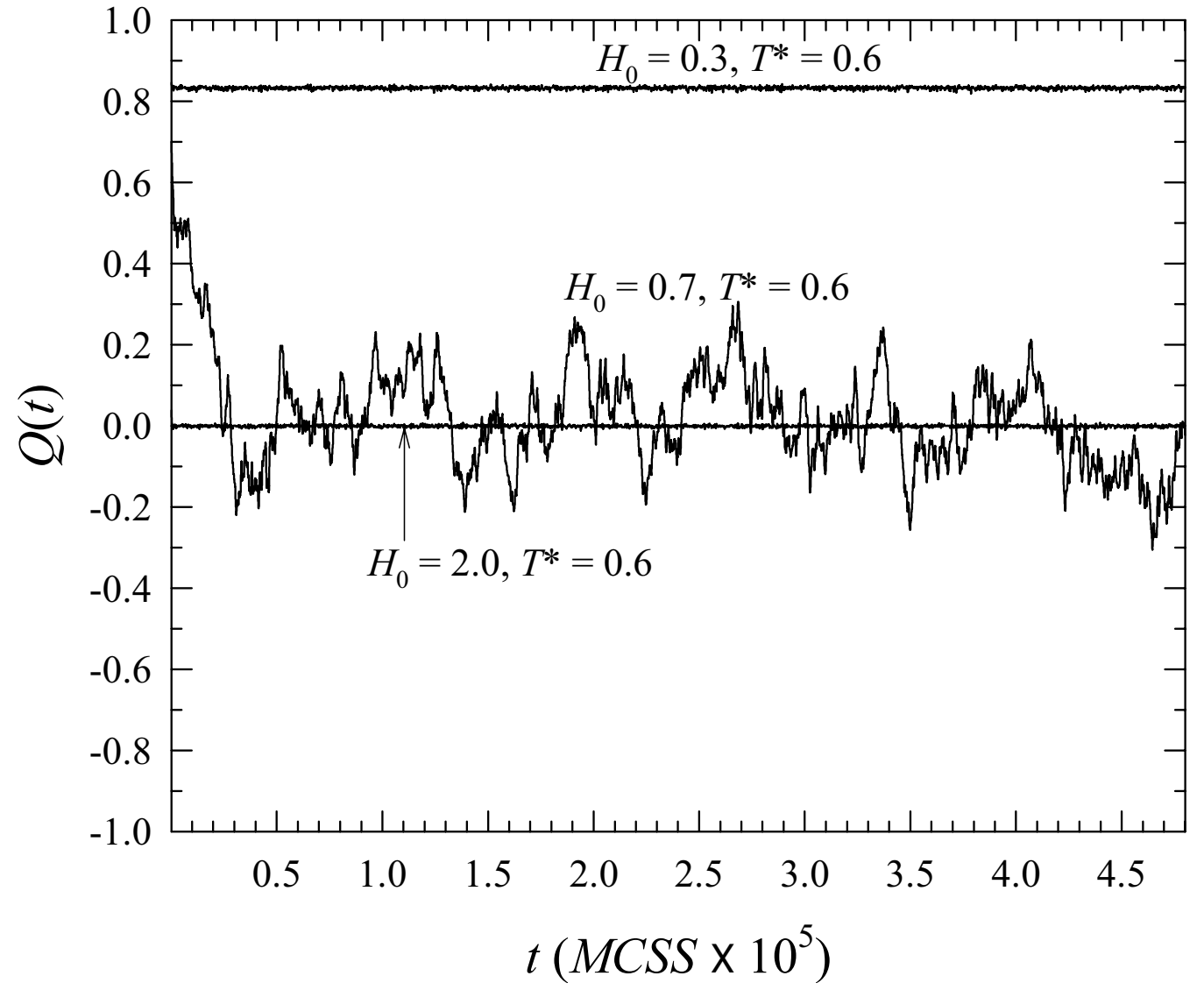

1.0
0.8
0.6
0.4
0.2
0.0
-0.2
-0.4
-0.6
-0.8
-1.0
Q(t)
H0 = 0.3, T* = 0.6
H0 = 0.7, T* = 0.6
H0 = 2.0, T* = 0.6
0.5 1.0 1.5 2.0 2.5 3.0 3.5 4.0 4.5
t (MCSS x 10^5)

Fig. 3

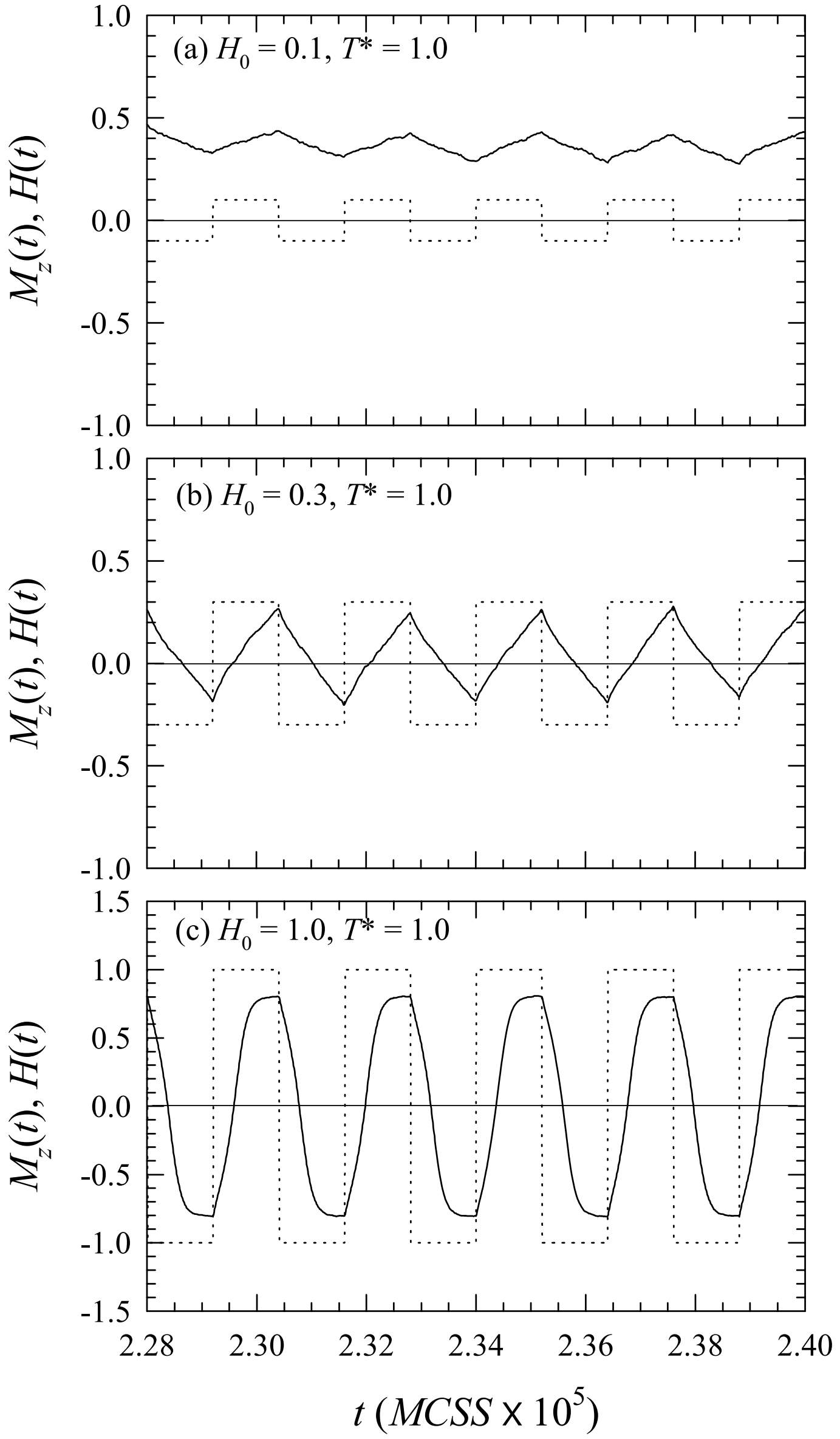

(a) $H_0$ = 0.1, $T^*$ = 1.0
(b) $H_0$ = 0.3, $T^*$ = 1.0
(c) $H_0$ = 1.0, $T^*$ = 1.0
$M_z(t)$, $H(t)$
1.5
1.0
0.5
0.0
-0.5
-1.0
-1.5
2.28
2.30
2.32
2.34
2.36
2.38
2.40
$t$ ($MCSS$ x $10^5$)

Fig. 4(a)

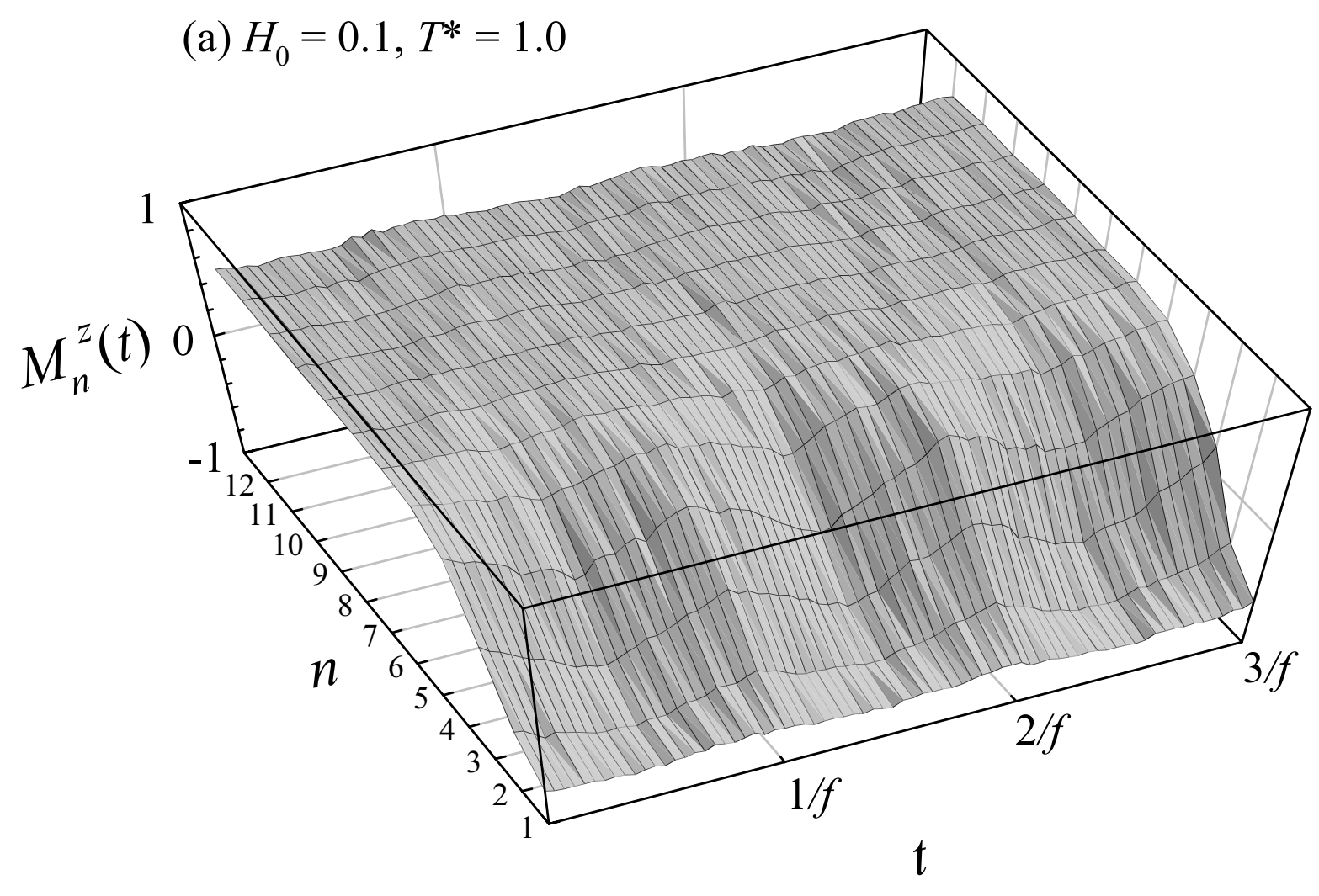

Fig. 4(b)

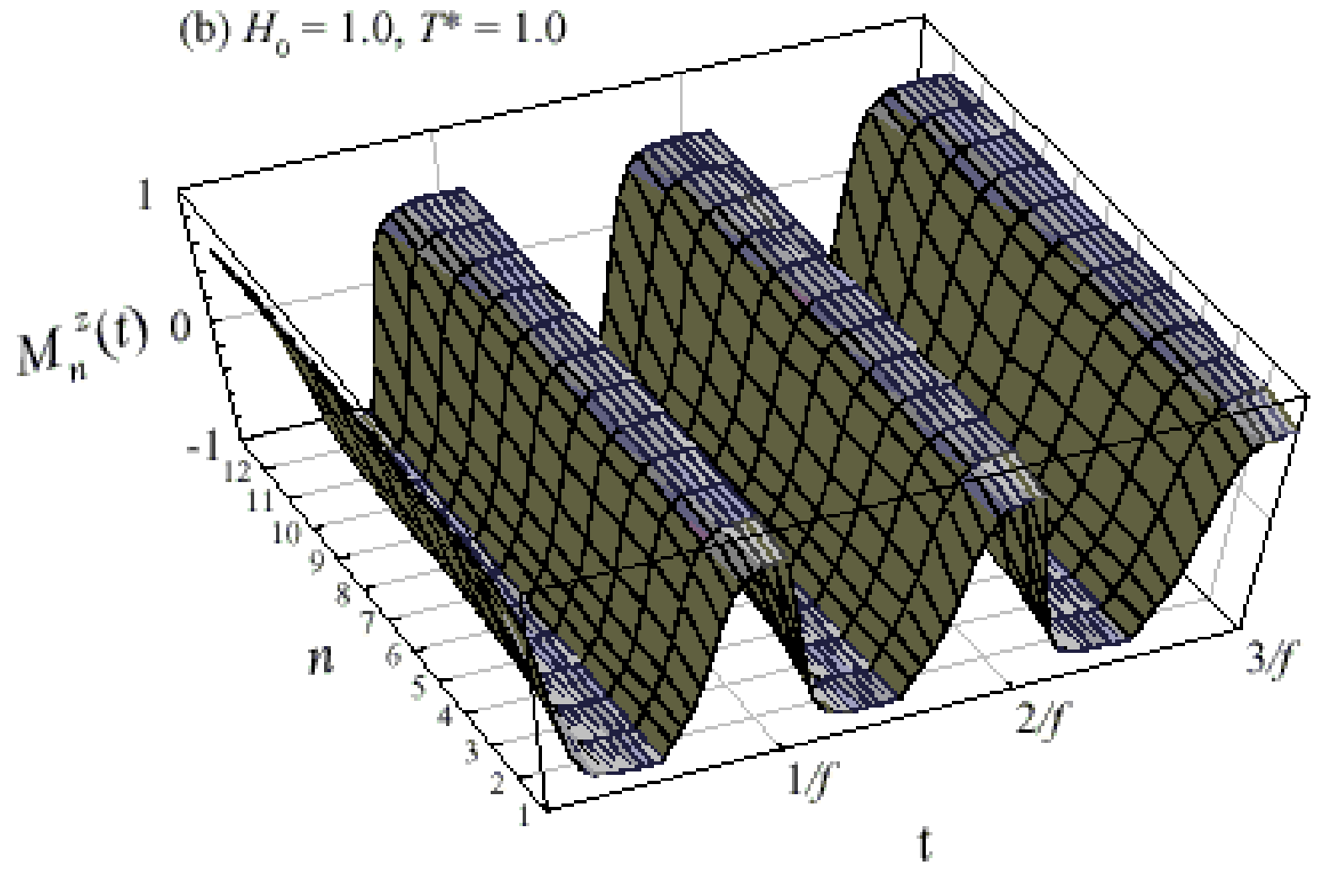

Fig. 5

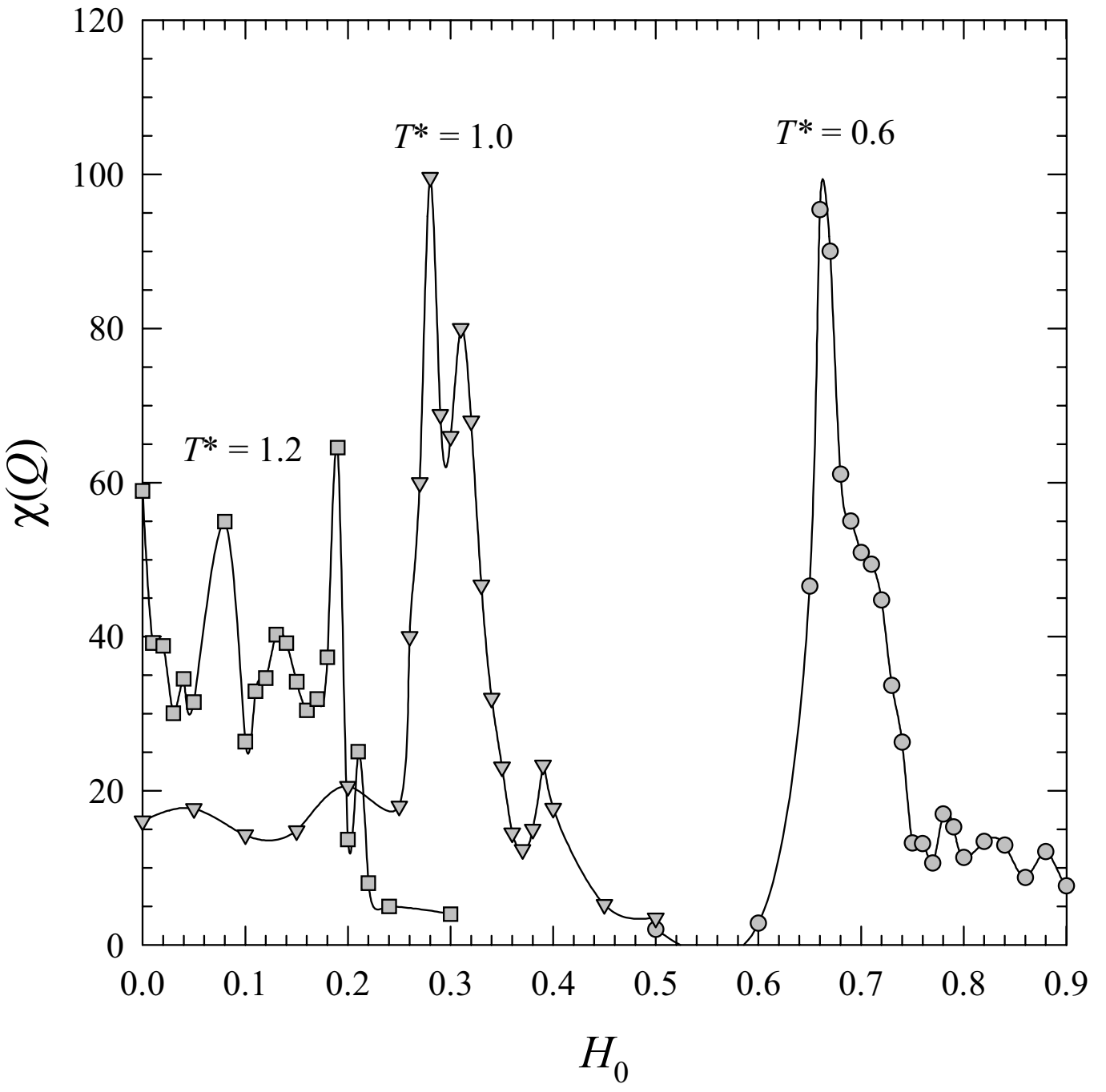

Fig. 6

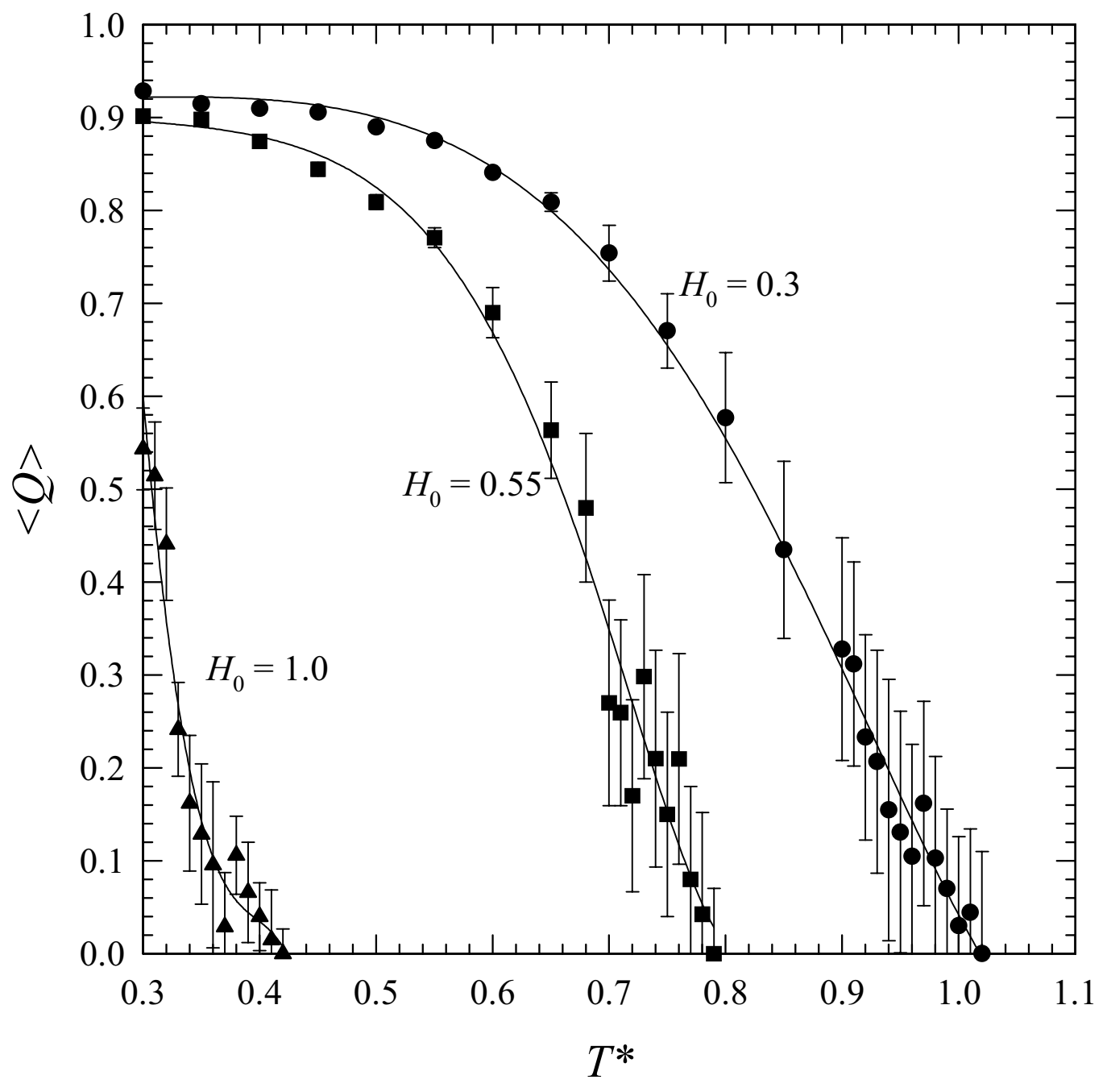

Fig. 7

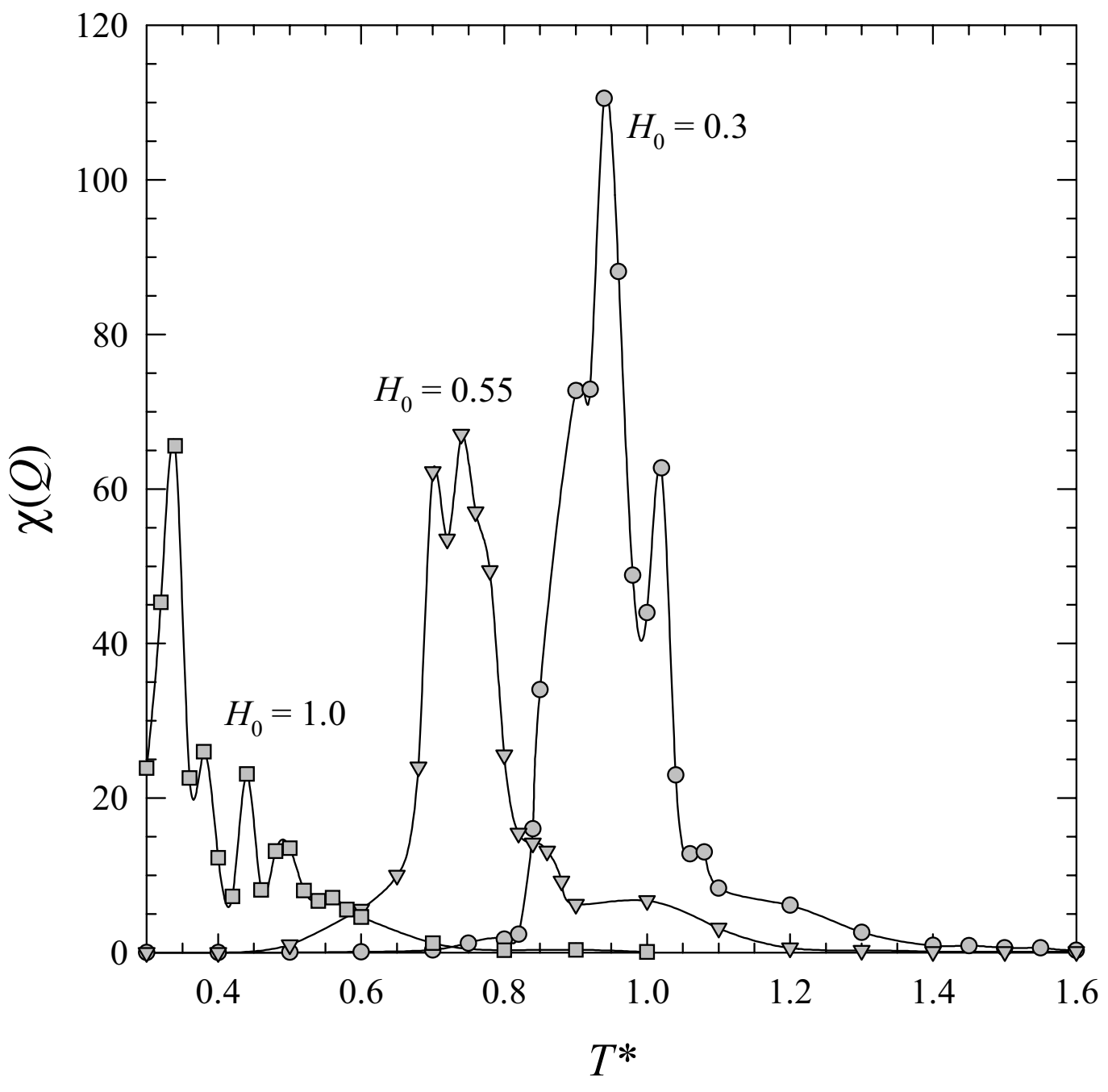

Fig. 8(a)

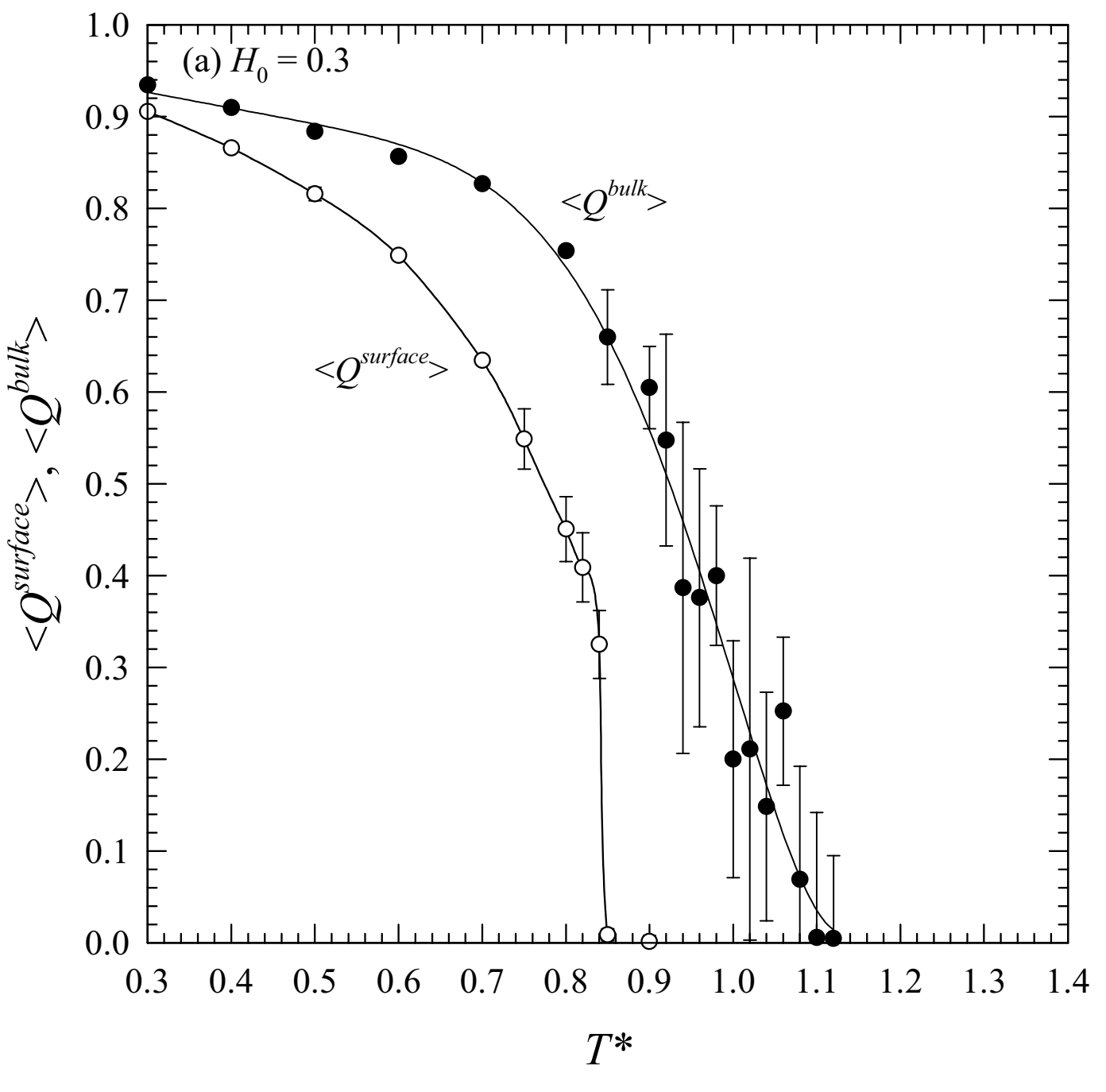

Fig. 8(b)

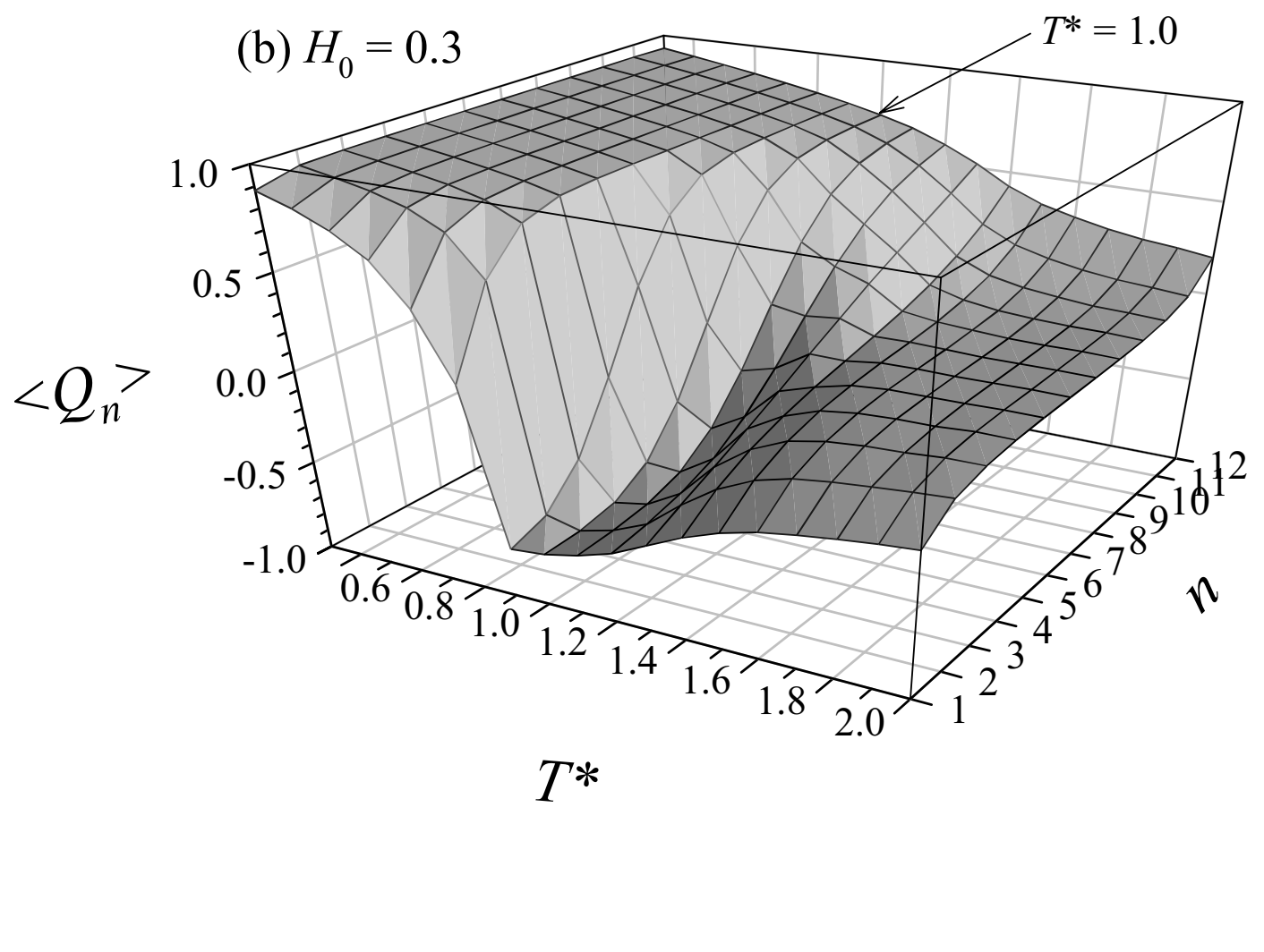

Fig. 9(a)(b)(c)

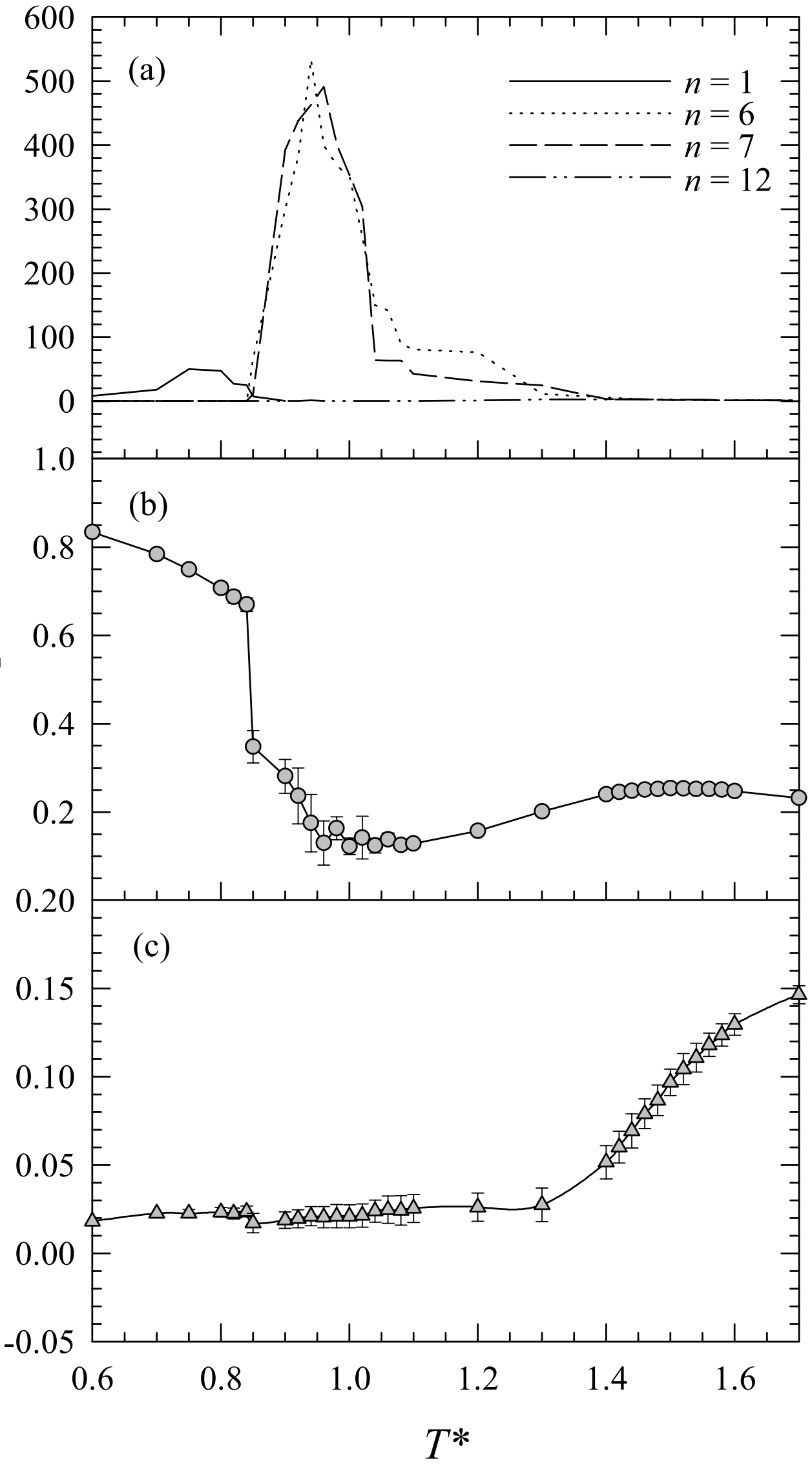

Fig.10

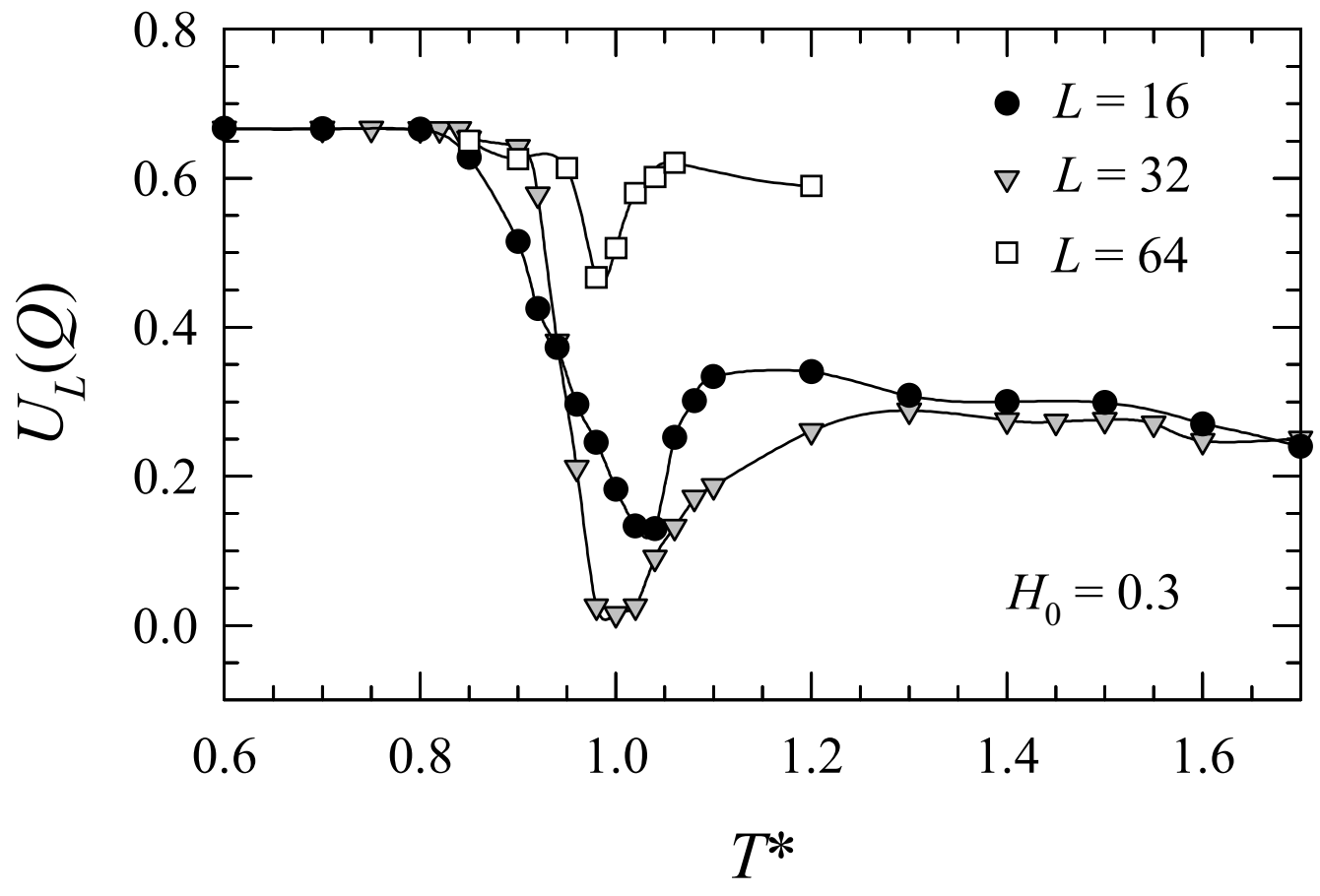

Fig. 11

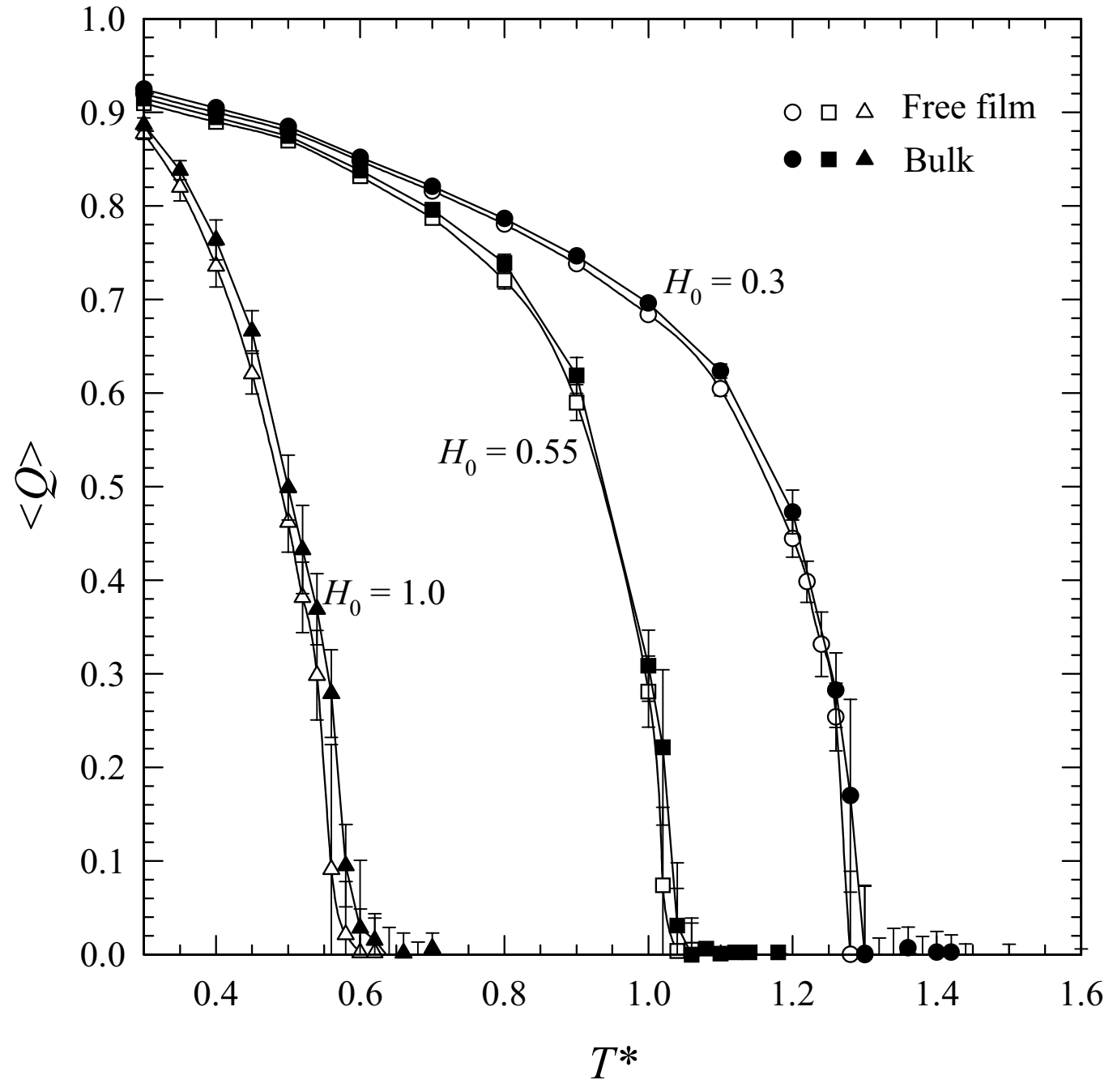

Fig. 12(a)(b)(c)

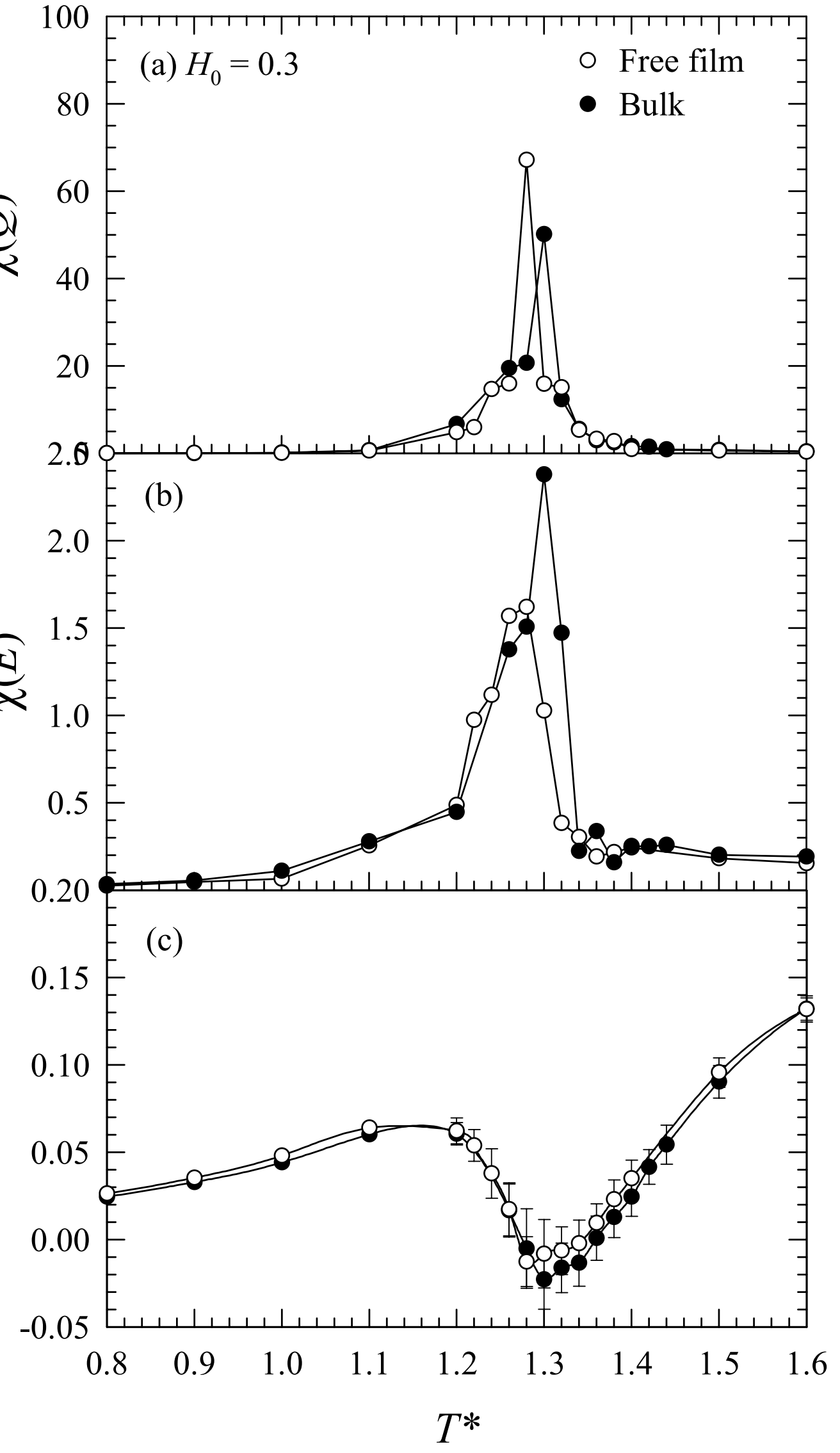

Fig. 13

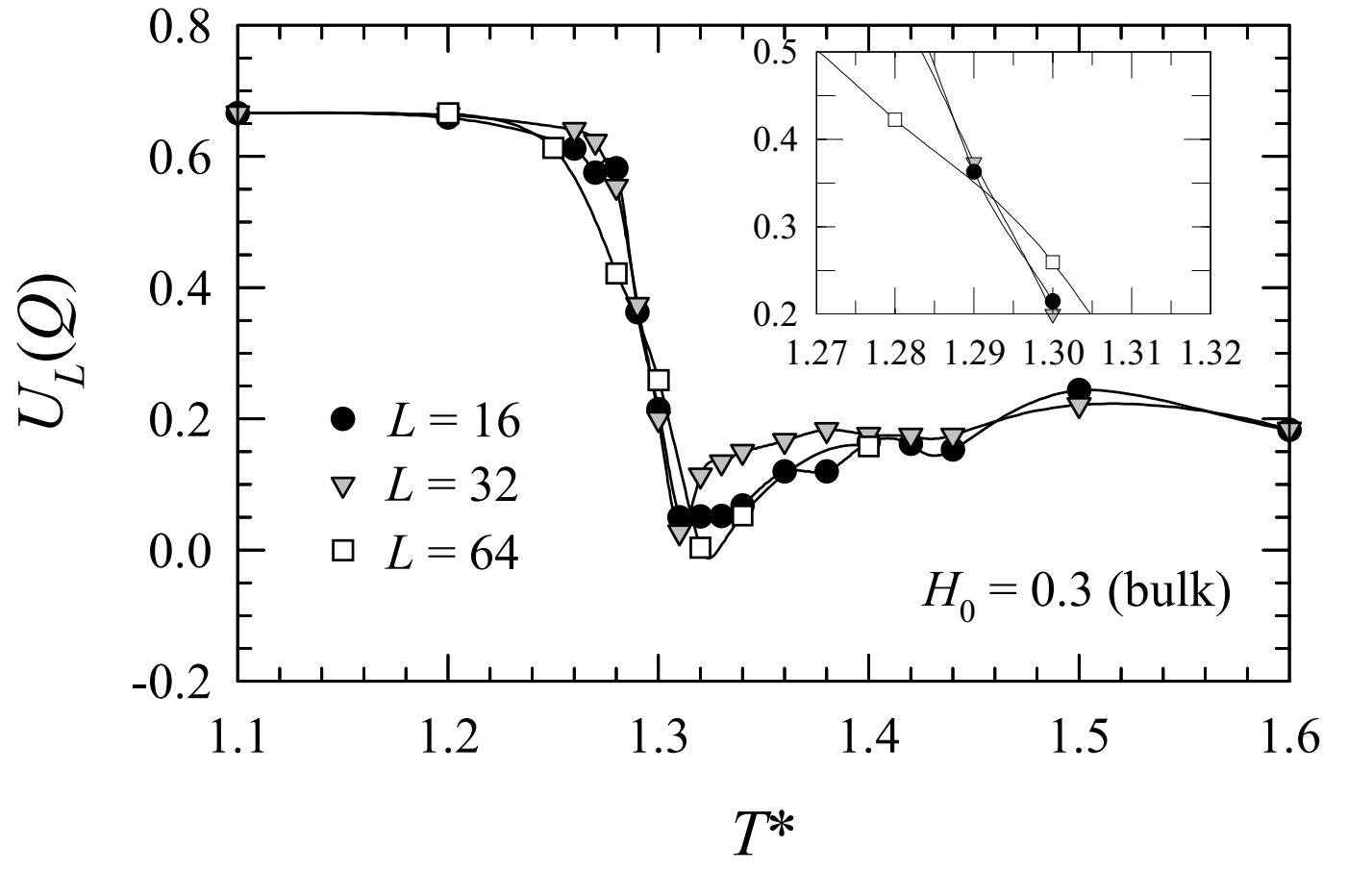

Fig. 14

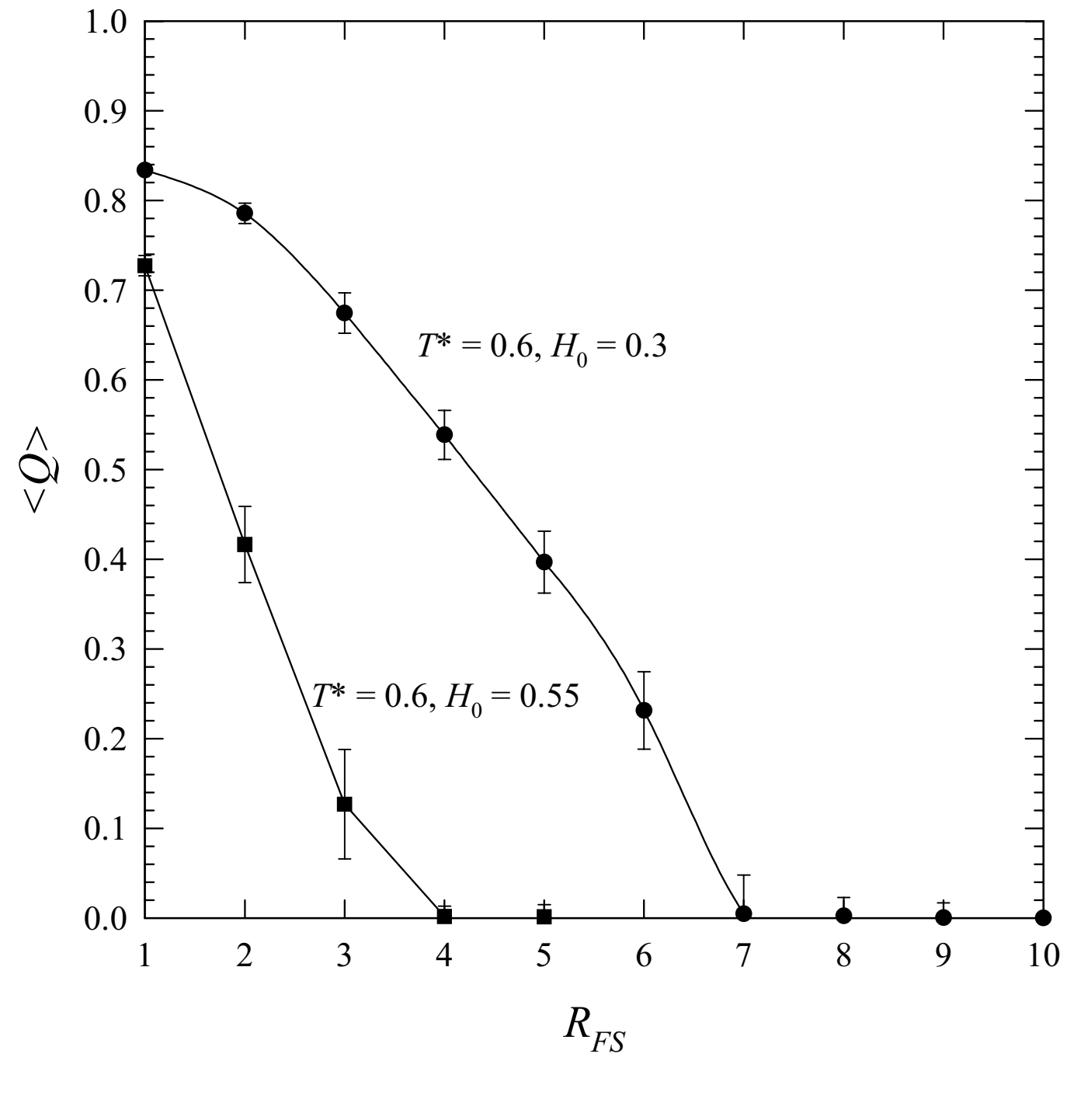

1.0
0.9
0.8
0.7
0.6
0.5
0.4
0.3
0.2
0.1
0.0
<$Q$>
$T$* = 0.6, $H_0$ = 0.3
$T$* = 0.6, $H_0$ = 0.55
1 2 3 4 5 6 7 8 9 10
$R_{FS}$

Fig. 15(a)(b)(c)

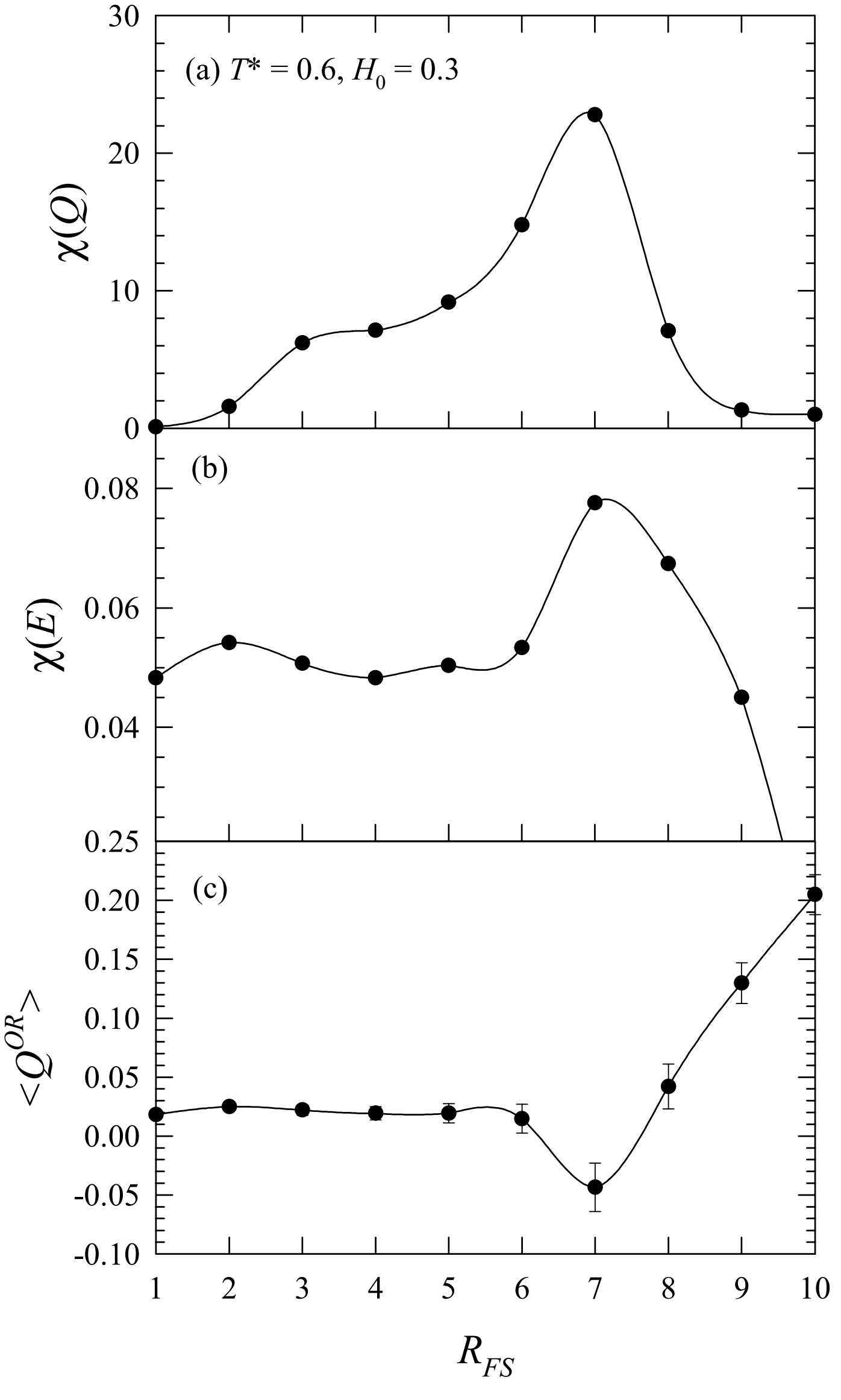